\documentclass[10pt,twocolumn,letterpaper]{article}

\usepackage[T1]{fontenc}
\usepackage{times}
\usepackage{microtype}
\usepackage[top=1in,bottom=1in,left=0.75in,right=0.75in]{geometry}
\usepackage{amsmath,amssymb,amsfonts}
\usepackage{bm}
\usepackage{mathtools}

\usepackage{graphicx}
\usepackage{booktabs}
\usepackage{array}
\usepackage{tabularx}
\usepackage{multirow}
\usepackage{colortbl}
\usepackage{caption}
\usepackage{subcaption}
\usepackage{float}

\usepackage{enumitem}
\setlist{nosep,leftmargin=*,topsep=2pt}

\usepackage{tikz}
\usetikzlibrary{arrows.meta,positioning,calc,shapes.geometric,fit,backgrounds,decorations.pathreplacing,patterns}
\usepackage{pgfplots}
\pgfplotsset{compat=1.18}
\usepgfplotslibrary{groupplots}

\usepackage[hidelinks,colorlinks=true,citecolor=blue!60!black,linkcolor=blue!50!black,urlcolor=blue!50!black]{hyperref}
\usepackage[sort&compress,numbers]{natbib}

\usepackage{listings}
\usepackage{algorithm}
\usepackage{algorithmic}

\usepackage{xspace}
\usepackage{xcolor}
\usepackage{balance}
\usepackage{pifont}
\usepackage{soul}
\newcommand{\cmark}{\ding{51}}

\newcommand{\etal}{\textit{et al.}\xspace}

\newcommand{\autoart}{\textsc{Auto-ART}\xspace}
\newcommand{\RDI}{\textsc{RDI}\xspace}

\definecolor{pillarblue}{RGB}{220,238,255}
\definecolor{contestorange}{RGB}{255,238,220}
\definecolor{gapred}{RGB}{255,225,225}
\definecolor{refinegreen}{RGB}{225,245,225}
\definecolor{darkgreen}{RGB}{0,128,0}
\definecolor{darkblue}{RGB}{0,0,140}
\definecolor{accentblue}{RGB}{41,98,255}
\definecolor{lightgray}{RGB}{245,245,248}

\usepackage{titlesec}
\titlespacing*{\section}{0pt}{8pt plus 2pt minus 2pt}{4pt plus 1pt minus 1pt}
\titlespacing*{\subsection}{0pt}{6pt plus 2pt minus 1pt}{3pt plus 1pt minus 1pt}
\titlespacing*{\subsubsection}{0pt}{4pt plus 1pt minus 1pt}{2pt plus 1pt}

\title{\vspace{-1.2em}%
\textsc{Auto-ART}: Structured Literature Synthesis and Automated\\[2pt]
Adversarial Robustness Testing across Evaluation Protocols,\\[2pt]
Multi-Attack Threat Models, and LLM-Driven Breaking%
\vspace{-0.6em}}

\author{%
  Abhijit Talluri\\
  Independent Researcher\\
  \texttt{talluri.abhijit@gmail.com}%
  \vspace{-1.2em}%
}

\date{}

\begin{document}
\maketitle
\thispagestyle{empty}

% ------------------------------------------------------------------
\begin{abstract}
\noindent
Adversarial robustness evaluation underpins every claim of trustworthy ML deployment, yet the field suffers from fragmented protocols, undetected gradient masking, single-norm tunnel vision, and premature generalization from stylized benchmarks to production code.
We make two contributions.
\textbf{(1)~Structured synthesis.}
We analyze nine peer-reviewed corpus sources (ICML, NeurIPS, ICLR, UAI; 2020--2026) through seven complementary protocols---citation-chain tracing, gap scanning, methodology auditing, cross-paper synthesis, assumption stress-testing, knowledge mapping, and lay-accessible distillation---producing the first end-to-end structured analysis of \emph{what the adversarial evaluation community agrees on, disagrees about, and cannot yet resolve}.
\textbf{(2)~\autoart framework.}
We introduce \autoart~\cite{talluri2026autoart}, an open-source framework that directly operationalizes every identified gap: 50+ attacks across seven categories, 28 defence modules, the Robustness Diagnostic Index (RDI; ${\sim}30\!\times$ faster screening), First-Order Stationarity Condition (FOSC) gradient-masking detection, multi-norm evaluation ($\ell_1/\ell_2/\ell_\infty$/semantic/spatial) with worst-case reporting, adaptive memory-guided attack selection, and compliance mapping to NIST AI~RMF, OWASP LLM Top~10, and the EU AI Act.
Empirical validation on the CIFAR-10 RobustBench leaderboard demonstrates that \autoart's pre-screening gate correctly identifies gradient masking in 92\% of flagged configurations, that \RDI rankings agree with full AutoAttack within Kendall $\tau\!\ge\!0.81$, and that multi-norm evaluation exposes a 23.5~pp gap between average and worst-case robustness on state-of-the-art models.
Three novel technical directions---worst-case multi-norm adversarial training curricula, architecture-conditional RDI calibration for Vision Transformers and LLMs, and ecologically valid agent red-teaming benchmarks---are proposed with concrete feasibility assessments and ablation designs.
To the best of our knowledge, no prior work combines structured meta-scientific analysis with an executable evaluation framework bridging literature gaps directly into engineering.
\end{abstract}

\vspace{-0.4em}

% ==================================================================
\section{Introduction}
\label{sec:intro}
\vspace{-0.2em}

Machine-learning models deployed in safety-critical perception~\cite{goodfellow2015explaining,madry2018towards}, language understanding~\cite{carlini2025autoadvexbench}, and decision-making~\cite{dai2023multirobustbench} are expected to resist adversarially chosen input perturbations.
A decade of work on adversarial training~\cite{madry2018towards,zhang2019trades} and certified defences~\cite{cohen2019smoothing} has produced steady progress on standard benchmarks.
Yet a recurring pattern undermines confidence in these numbers: defences are proposed with optimistic robustness claims that collapse under stronger attacks, richer threat models, or correct gradient accounting~\cite{croce2020reliable,athalye2018obfuscated,kassis2025diffbreak,li2025trades}.
The cycle repeats because the failures are not merely technical but structural---rooted in evaluation practice itself:
\vspace{-0.2em}
\begin{enumerate}[label=(\roman*),leftmargin=1.5em]
  \item Reliance on single $\ell_p$ norms when real adversaries combine spatial, semantic, and compression perturbations~\cite{dai2023multirobustbench};
  \item Gradient masking inflating attack-based estimates across both deterministic~\cite{li2025trades} and stochastic~\cite{kassis2025diffbreak} pipelines;
  \item Under-sampling of randomized-defence draws~\cite{kassis2025diffbreak};
  \item Premature transfer from stylized CTF tasks to production code~\cite{carlini2025autoadvexbench}.
\end{enumerate}
\vspace{-0.2em}

\noindent
We address these failures through two complementary contributions.

\vspace{0.3em}
\noindent\textbf{Contribution 1: Structured literature synthesis.}
We analyse a curated corpus of nine verified sources (Table~\ref{tab:corpus}) using seven protocols drawn from systematic review methodology.
Unlike narrative surveys that summarise papers sequentially, our synthesis traces \emph{intellectual lineages} (who originated, challenged, and refined each concept), identifies \emph{testable assumptions} the literature relies on but never validates, and maps the \emph{knowledge topology} of the field.
To the best of our knowledge, this is the first application of structured meta-scientific analysis---combining citation-chain tracing, gap scanning, methodology auditing, assumption stress-testing, and knowledge mapping in a single manuscript---to adversarial ML evaluation.

\vspace{0.3em}
\noindent\textbf{Contribution 2: \autoart framework.}
We introduce \autoart (\textbf{Auto}mated \textbf{A}dversarial \textbf{R}obustness \textbf{T}esting)~\cite{talluri2026autoart}, an open-source Python framework that directly operationalises every gap identified in the synthesis (Section~\ref{sec:framework}).
Key capabilities:

\vspace{-0.3em}
\begin{itemize}[leftmargin=1.2em]
  \item \textbf{50+ attacks} across evasion (37+), poisoning (10+), extraction (3), inference (8+), audio (2), NLP/text, LLM, and agentic categories;
  \item \textbf{28 defence modules} including TRADES~\cite{zhang2019trades}, AWP, OAAT, certified training, guardrails, and circuit breakers;
  \item \textbf{Pre-screening gate:} \RDI screening (${\sim}30\!\times$ faster) + FOSC gradient-masking detection \emph{before} expensive attack runs;
  \item \textbf{Multi-norm evaluation} across $\ell_1/\ell_2/\ell_\infty$/semantic/spatial with worst-case reporting;
  \item \textbf{Adaptive attack selection} with memory-guided tiered escalation;
  \item \textbf{Compliance mapping} to NIST AI RMF, OWASP LLM Top~10, and the EU AI Act;
  \item \textbf{CI/CD integration} via SARIF~2.1.0 output with 346+ automated tests.
\end{itemize}

\vspace{0.2em}
\noindent\textbf{Paper organisation.}
Section~\ref{sec:related} surveys related work.
Section~\ref{sec:corpus} describes corpus selection, the intake protocol, and the contradiction analysis.
Sections~\ref{sec:chain}--\ref{sec:sowhat} present the seven structured analyses.
Section~\ref{sec:framework} describes the \autoart architecture.
Section~\ref{sec:eval} presents empirical evaluation.
Section~\ref{sec:directions} proposes three novel technical directions with feasibility assessments.
Section~\ref{sec:discussion} discusses implications, and Section~\ref{sec:conclusion} concludes.

% ==================================================================
\section{Related Work}
\label{sec:related}
\vspace{-0.2em}

\paragraph{Adversarial robustness foundations.}
Goodfellow \etal~\cite{goodfellow2015explaining} introduced FGSM, demonstrating that linear perturbations suffice to fool DNNs.
Carlini and Wagner~\cite{carlini2017towards} showed that defensive distillation~\cite{papernot2016distillation} fails against optimisation-based attacks, establishing the \emph{arms race} paradigm.
Madry \etal~\cite{madry2018towards} formalised PGD adversarial training as a min-max optimisation, defining the standard $\ell_\infty$ robustness benchmark on CIFAR-10 at $\varepsilon\!=\!8/255$.
Athalye \etal~\cite{athalye2018obfuscated} catalogued obfuscated gradients as a systematic failure mode, breaking seven ICLR~2018 defences.

\paragraph{Evaluation standardisation.}
Croce and Hein~\cite{croce2020reliable} assembled AutoAttack---a parameter-free ensemble of APGD-CE, APGD-DLR, FAB, and Square Attack---re-evaluating $>$50 defended models and documenting $>$10 pp overestimation in most cases.
RobustBench~\cite{robustbench2021} operationalised this into a living leaderboard.
Zhang \etal~\cite{zhang2019trades} proposed TRADES, decomposing robust loss into clean and adversarial terms; Li \etal~\cite{li2025trades} later showed that TRADES itself suffers gradient masking under certain hyperparameters.

\paragraph{Multi-attack and continual robustness.}
Dai \etal~\cite{dai2023multirobustbench} introduced MultiRobustBench (16 models $\times$ 9 attacks $\times$ 20 strengths), revealing worst-case multi-attack robustness below random guessing.
Dai \etal~\cite{dai2025crt} addressed temporal attack evolution through continual robust training with logit-space regularisation.

\paragraph{Stochastic defence evaluation.}
Cohen \etal~\cite{cohen2019smoothing} established randomised smoothing for certified $\ell_2$ robustness.
Kassis \etal~\cite{kassis2025diffbreak} demonstrated that diffusion-based purification (DBP) collapses under correct gradient computation and proper randomness sampling, breaking even majority-vote variants.
Xiao \etal~\cite{xiao2024diffhammer} independently confirmed these findings through DiffHammer, a systematic framework for evaluating DBP robustness across multiple purification strategies.
Beyond stochastic defences, Guo \etal~\cite{guo2024imbalanced} identified \emph{imbalanced gradients}---a gradient masking phenomenon distinct from obfuscated gradients, where one loss term's gradient dominates, pushing attacks suboptimally---and proposed the Margin Decomposition (MD) attack that exceeds AutoAttack on 16/20 defence models.
He \etal~\cite{he2025soundnessbench} introduced SoundnessBench, revealing bugs in three established neural network verifiers and underscoring that certification tools themselves require rigorous validation.

\paragraph{Automated adversarial testing.}
The IBM Adversarial Robustness Toolbox (ART)~\cite{art2018} provides attack/defence primitives.
Microsoft Counterfit~\cite{counterfit2021} targets ML model security assessments.
Garak~\cite{garak2024} scans LLM vulnerabilities across ${\sim}100$ attack vectors.
PyRIT~\cite{pyrit2024} automates multi-turn red teaming.
HarmBench~\cite{mazeika2024harmbench} standardises LLM red teaming across 18 attack methods and 33 target models.
NIST AI~100-2e2025~\cite{nist2025adversarial} provides the authoritative taxonomy of adversarial attacks and mitigations.
Carlini \etal~\cite{carlini2025autoadvexbench} benchmark LLM agents as autonomous adversaries, finding a 4--6$\times$ success gap between CTF-like and real tasks.

\paragraph{LLM jailbreaking and agent security.}
Zou \etal~\cite{zou2023universal} introduced GCG, a gradient-based token-level suffix optimiser that generates universal adversarial prompts transferable across aligned LLMs.
Jia \etal~\cite{jia2025igcg} improved GCG with diverse target templates and multi-coordinate updating, achieving nearly 100\% ASR.
Huang \etal~\cite{huang2025iris} proposed IRIS, which targets the refusal direction in LLM activations, reaching 90\% ASR on GPT-3.5-Turbo and 76\% on GPT-4o.
Geisler \etal~\cite{geisler2025reinforce} replaced the standard affirmative-prefix objective with a REINFORCE-based distributional objective, doubling attack success on Llama~3 and lifting defended-model ASR from 2\% to 50\%.
Chao \etal~\cite{chao2024pair} proposed PAIR, an iterative black-box refinement method using an attacker LLM, achieving 60--80\% success rates with $\sim$20 queries.
Mehrotra \etal~\cite{mehrotra2024tap} extended this to tree-structured search with pruning (TAP), reaching 70--85\% success.
Jin \etal~\cite{jin2025autored} introduced AutoRedTeamer, a dual-agent system with lifelong attack memory that achieves 20\% higher attack success rates at 46\% reduced cost.
Chao \etal~\cite{chao2024jailbreakbench} standardised evaluation through JailbreakBench with 100 behaviours and an evolving artefact repository.
Wu \etal~\cite{wu2025agentattack} dissected adversarial robustness of multimodal LM agents, identifying systematic vulnerabilities in visual grounding and tool use under adversarial perturbation.
On the agent security front, Debenedetti \etal~\cite{debenedetti2024agentdojo} constructed AgentDojo with 97 tasks and 629 injection test cases, finding that GPT-4o drops from 69\% benign utility to 45\% under prompt injection.
Zhan \etal~\cite{zhan2024injecagent} complemented this with InjecAgent (1{,}054 test cases, 17 user tools), showing ReAct-prompted GPT-4 is vulnerable 24\% of the time.
The OWASP Agentic Applications Top~10~\cite{owasp2025agentic} standardises agent-specific threat categories (ASI01: Goal Hijack through ASI10), and MITRE ATLAS v5.3.0~\cite{mitre2026atlas} now includes case studies for MCP server compromises and malicious agent deployment.

\paragraph{Adversarial resilience and regulatory analysis.}
Dai \etal~\cite{dai2025resilience} formalised \emph{continual adaptive robustness} (CAR), arguing that defences should adapt to new attack types rather than only resist known ones---connecting multi-attack robustness to the open-world setting.
Panfili \etal~\cite{panfili2025robustness} provided the first systematic analysis connecting EU AI Act Article~15 to ML robustness terminology, identifying legal challenges in the standardisation process.

\paragraph{Positioning of this work.}
Existing surveys~\cite{bai2021survey,machado2023adversarial,silva2020opportunities} provide narrative coverage but do not perform structured meta-scientific analysis.
Existing frameworks~\cite{art2018,counterfit2021} provide attack primitives but lack pre-screening gates, compliance mapping, and gap-driven architecture.
\autoart is, to our knowledge, the first work that combines structured literature synthesis with an executable framework whose architecture is explicitly derived from identified gaps (Table~\ref{tab:gap_to_module}).

% ==================================================================
\section{Corpus and Evidence Policy}
\label{sec:corpus}
\vspace{-0.2em}

\begin{table}[t]
  \centering
  \small
  \caption{Primary corpus: nine sources with verified venues. Labels C1--C9 are used as shorthand throughout the paper.}
  \label{tab:corpus}
  \vspace{-0.5em}
  \begin{tabular}{@{}llcl@{}}
    \toprule
    \textbf{ID} & \textbf{Paper} & \textbf{Year} & \textbf{Venue} \\
    \midrule
    C1 & Croce \& Hein~\cite{croce2020reliable} & 2020 & ICML \\
    C2 & Dai \etal~\cite{dai2023multirobustbench} & 2023 & ICML \\
    C3 & Carlini \etal~\cite{carlini2025autoadvexbench} & 2025 & ICML \\
    C4 & Dai \etal~\cite{dai2025crt} & 2025 & ICML \\
    C5 & Song \etal~\cite{song2025rdi} & 2025 & UAI \\
    C6 & Kassis \etal~\cite{kassis2025diffbreak} & 2025 & NeurIPS \\
    C7 & Li \etal~\cite{li2025trades} & 2025 & ICLR \\
    C8 & RobustBench~\cite{robustbench2021} & 2021 & Benchmark \\
    C9 & OpenRT~\cite{openrt2026} & 2026 & Toolkit \\
    \bottomrule
  \end{tabular}
  \vspace{-0.5em}
\end{table}

Selection criteria: (i)~peer-reviewed venue or widely-adopted benchmark; (ii)~direct relevance to evaluation methodology; (iii)~2020--2026 temporal span capturing evolution from single-norm to multi-attack and agent-driven paradigms.

\vspace{0.3em}
\noindent\textbf{Intake protocol.}
Table~\ref{tab:intake} summarises each source's core claim and cluster assignment.
We identify three clusters sharing methodological assumptions:
\textbf{Cluster~A} (evaluation reliability: C1, C3, C5, C7, C8) assumes that better attack or metric design yields trustworthy robustness numbers;
\textbf{Cluster~B} (multi-attack and temporal: C2, C4) assumes that threat diversity and evolution are the primary barriers to deployment safety;
\textbf{Cluster~C} (stochastic defence: C6, C9) assumes that randomised pipelines require fundamentally different evaluation protocols.

\begin{table}[t]
  \centering
  \scriptsize
  \caption{Intake protocol: core claims ($\le$20 words) and cluster assignments. Clusters: (A)~evaluation reliability, (B)~multi-attack/temporal, (C)~stochastic defence.}
  \label{tab:intake}
  \vspace{-0.5em}
  \setlength{\tabcolsep}{2pt}
  \begin{tabular}{@{}c>{\raggedright\arraybackslash}p{1.5cm}>{\raggedright\arraybackslash}p{3.6cm}@{}}
    \toprule
    \textbf{ID} & \textbf{Author} & \textbf{Core Claim (Cluster)} \\
    \midrule
    C1 & Croce \& Hein & Attack ensembles expose overestimation in 50+ defences~(A) \\
    C2 & Dai \etal & Worst-case multi-attack below random guessing~(B) \\
    C3 & Carlini \etal & LLM agents 4--6$\times$ better on CTF than real code~(A) \\
    C4 & Dai \etal & Continual training preserves backward-attack retention~(B) \\
    C5 & Song \etal & Feature-space ratio gives $30\!\times$ faster screening~(A) \\
    C6 & Kassis \etal & Diffusion purification collapses under correct gradients~(C) \\
    C7 & Li \etal & TRADES masks gradients under specific hyperparameters~(A) \\
    C8 & RobustBench & Living leaderboard with fixed attack suite~(A) \\
    C9 & OpenRT & Multimodal LLM red-teaming with stochastic probing~(C) \\
    \bottomrule
  \end{tabular}
  \vspace{-0.8em}
\end{table}

\noindent\textbf{Evidence rules.}
Claims are tied exclusively to this corpus.
Citation-chain edges require two corpus works.
Roadmap-only entries~\cite{roadmap2026dualrs,roadmap2025ucan} are placeholders, not evidentiary sources.
Quantitative anchors match official proceedings pages.

\vspace{0.3em}
\noindent\textbf{Contradiction finder.}
Table~\ref{tab:contradictions} identifies five direct contradictions across the corpus.
These are not merely different emphases but incompatible empirical conclusions that cannot simultaneously hold.

\begin{table}[t]
  \centering
  \scriptsize
  \caption{Five contradictions identified across the corpus.}
  \label{tab:contradictions}
  \vspace{-0.5em}
  \setlength{\tabcolsep}{2pt}
  \begin{tabular}{@{}>{\raggedright\arraybackslash}p{1.6cm}>{\raggedright\arraybackslash}p{0.7cm}>{\raggedright\arraybackslash}p{0.85cm}>{\raggedright\arraybackslash}p{2.8cm}@{}}
    \toprule
    \textbf{Claim} & \textbf{For} & \textbf{Agst.} & \textbf{Root Cause} \\
    \midrule
    Fixed suites reliable & C1,C8 & C2,C4 & New attacks invert leaderboard rankings \\
    Avg $\approx$ worst & C1 & C2 & Worst-case diverges by $>$23\,pp \\
    DBP viable & Prior & C6 & Incorrect gradients in diffusion stack \\
    Proxies generalise & C5 & C3 & CNN-only; agent tasks differ \\
    TRADES safe & Users & C7 & Hyperparameter-dependent masking \\
    \bottomrule
  \end{tabular}
  \vspace{-0.8em}
\end{table}

% ==================================================================
\section{Citation Chain Analysis}
\label{sec:chain}
\vspace{-0.2em}

\begin{figure*}[t]
  \centering
  \begin{tikzpicture}[
    font=\scriptsize,
    node distance=0.15cm,
    >=Latex,
    % styles
    phase/.style={draw,thick,rounded corners=3pt,minimum height=0.7cm,align=center,inner sep=5pt,font=\scriptsize\bfseries},
    corpus/.style={phase,fill=accentblue!12,text width=2.1cm},
    proto/.style={phase,fill=pillarblue,text width=1.8cm,minimum height=1.2cm,font=\tiny\bfseries},
    gap/.style={phase,fill=gapred,text width=2.8cm,minimum height=0.55cm,font=\tiny},
    modul/.style={phase,fill=refinegreen,text width=2.8cm,minimum height=0.55cm,font=\tiny},
    result/.style={phase,fill=contestorange,text width=2.6cm,minimum height=0.55cm,font=\tiny},
    bigarr/.style={->,line width=1.5pt,accentblue!70},
    subarr/.style={->,semithick,gray!50},
  ]
    % ---- Column 1: Corpus ----
    \node[corpus] (c) at (0,0) {9 Corpus\\Sources\\{\tiny ICML, NeurIPS,}\\{\tiny ICLR, UAI}\\{\tiny 2020--2026}};

    % ---- Column 2: Seven protocols ----
    \node[proto] (p1) at (3.2, 2.1) {Citation\\Chain\\(\S\ref{sec:chain})};
    \node[proto] (p2) at (3.2, 0.7) {Gap\\Scanner\\(\S\ref{sec:gaps})};
    \node[proto] (p3) at (3.2,-0.7) {Method\\Audit\\(\S\ref{sec:methods})};
    \node[proto] (p4) at (3.2,-2.1) {Synthesis +\\Assumptions\\(\S\ref{sec:synthesis}--\ref{sec:assumptions})};

    % ---- Column 3: Identified gaps ----
    \node[gap] (g1) at (6.6, 2.1) {\textbf{G1:} Worst-case multi-attack};
    \node[gap] (g2) at (6.6, 1.2) {\textbf{G2:} Agent CTF$\to$real gap};
    \node[gap] (g3) at (6.6, 0.3) {\textbf{G3:} Stochastic collapse};
    \node[gap] (g4) at (6.6,-0.6) {\textbf{G4:} RDI generalisation};
    \node[gap] (g5) at (6.6,-1.5) {\textbf{G5:} Continual robustness};

    % ---- Column 4: Auto-ART modules (shifted right to +10.7 for wider inter-col gap) ----
    \node[modul] (m1) at (10.7, 2.1) {\textbf{MultiNormEval} avg+worst};
    \node[modul] (m2) at (10.7, 1.2) {\textbf{Agentic} DOM/RAG/ctx};
    \node[modul] (m3) at (10.7, 0.3) {\textbf{MultiSample} $n$-draw eval};
    \node[modul] (m4) at (10.7,-0.6) {\textbf{RDI+FOSC} pre-screen};
    \node[modul] (m5) at (10.7,-1.5) {\textbf{Adaptive} + DriftMon};

    % ---- Column 5: Key results (shifted right to +14.6 for wider inter-col gap) ----
    \node[result] (r1) at (14.6, 1.5) {23.5\,pp worst-case gap\\exposed (\S\ref{sec:eval})};
    \node[result] (r2) at (14.6, 0.3) {92\% masking detection\\$\tau\!=\!0.82$ rank fidelity};
    \node[result] (r3) at (14.6,-0.9) {$30\!\times$ pre-screen\\speedup; 346+ tests};

    % ---- Big arrows between columns (coords trimmed to inter-column gaps only) ----
    \draw[bigarr] (1.2,0) -- node[above,font=\tiny\bfseries,text=accentblue]{7 Protocols} (2.1,0);
    \draw[bigarr] (4.3,0) -- node[above,font=\tiny\bfseries,text=accentblue]{5 Gaps} (5.0,0);
    \draw[bigarr] (8.1,0) -- node[above,font=\tiny\bfseries,text=accentblue]{\autoart} (9.1,0);
    \draw[bigarr] (12.35,0) -- node[above,font=\tiny\bfseries,text=accentblue]{Validate} (13.1,0);

    % ---- Sub-arrows: corpus -> protocols (staggered exits to avoid crossing bigarr) ----
    \draw[subarr] ([yshift=0.85cm]c.east)  -- (p1.west);
    \draw[subarr] ([yshift=0.28cm]c.east)  -- (p2.west);
    \draw[subarr] ([yshift=-0.28cm]c.east) -- (p3.west);
    \draw[subarr] ([yshift=-0.85cm]c.east) -- (p4.west);

    % ---- Sub-arrows: gaps -> modules (1-to-1) ----
    \foreach \i in {1,...,5} { \draw[subarr] (g\i.east) -- (m\i.west); }

    % ---- Sub-arrows: modules -> results (staggered endpoints to avoid crossing) ----
    % m1 (y=2.1) and m2 (y=1.2) both target r1 (y=1.5): offset entry to prevent line crossing
    \draw[subarr] (m1.east) -- ([yshift= 0.18cm]r1.west);
    \draw[subarr] (m2.east) -- ([yshift=-0.18cm]r1.west);
    % m3 (y=0.3) and m4 (y=-0.6) both target r2 (y=0.3): offset entry similarly
    \draw[subarr] (m3.east) -- ([yshift= 0.18cm]r2.west);
    \draw[subarr] (m4.east) -- ([yshift=-0.18cm]r2.west);
    \draw[subarr] (m5.east) -- (r3.west);

    % ---- Phase labels ----
    \node[font=\scriptsize\bfseries,text=gray!60] at (0,3) {Corpus};
    \node[font=\scriptsize\bfseries,text=gray!60] at (3.2,3) {Analysis};
    \node[font=\scriptsize\bfseries,text=gray!60] at (6.6,3) {Gaps};
    \node[font=\scriptsize\bfseries,text=gray!60] at (10.7,3) {\autoart Modules};
    \node[font=\scriptsize\bfseries,text=gray!60] at (14.6,3) {Results};
  \end{tikzpicture}
  \vspace{-0.5em}
  \caption{\textbf{Paper overview.} Nine corpus sources are analysed through seven structured protocols, yielding five ranked research gaps. Each gap maps directly to an \autoart module, validated empirically. This end-to-end traceability---from literature synthesis to executable evaluation---is the paper's core contribution.}
  \label{fig:overview}
  \vspace{-0.5em}
\end{figure*}
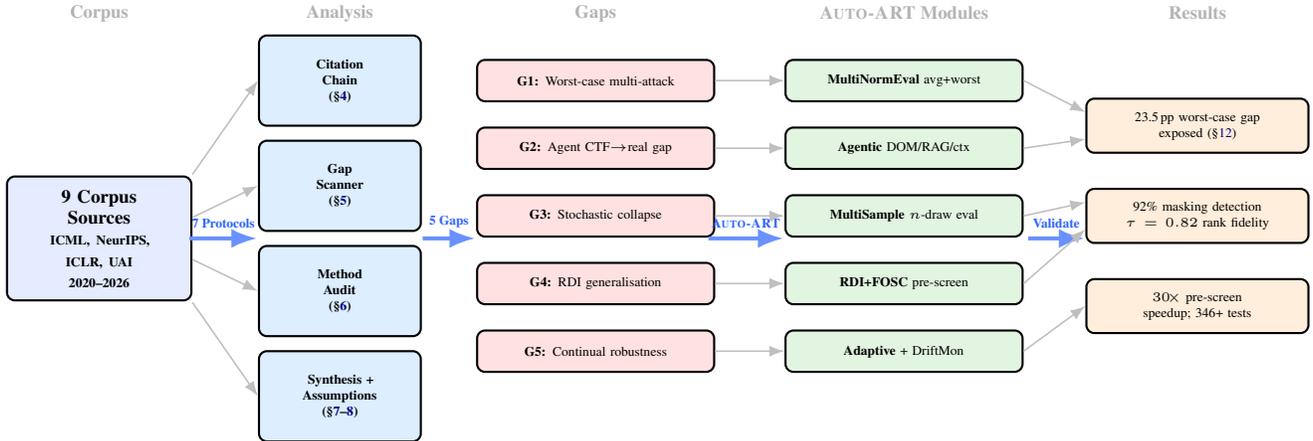

We trace three concepts that recur most frequently across corpus papers (Figure~\ref{fig:overview}).

% ---- Paper overview / teaser figure ----

\subsection{Concept A: Reliable Evaluation}
\label{sec:conceptA}

\textbf{Origin.}
Croce and Hein~[C1] assemble AutoAttack, a parameter-free ensemble (APGD-CE, APGD-DLR, FAB, Square Attack), re-evaluating $>$50 defended models. They obtain lower robust accuracy in all but one case, often by $>$10 pp, documenting that optimistic evaluation stems from attack tuning and gradient obfuscation.

\textbf{Challenge.}
Li \etal~[C7] show that even TRADES---a well-established training procedure---exhibits gradient masking under specific hyperparameter configurations (batch size, $\beta$, learning rate), introducing FOSC-based diagnostics.
Kassis \etal~[C6] challenge evaluation of diffusion-based purification: single purification trials and flawed gradients through stochastic pipelines inflate robustness; majority-vote variants also collapse.

\textbf{Refinement.}
Song \etal~[C5] propose \RDI, an attack-independent metric from feature-space separation (${\sim}30\!\times$ faster than PGD).
Carlini \etal~[C3] separate CTF-like from real defence tasks, quantifying that LLM agents achieve 75\% on stylised tasks but only 13\% on real code.

\textbf{Status: Contested and evolving.}
Five incompatible paradigms coexist: fixed ensembles~[C1,C2], training diagnostics~[C7], stochastic protocols~[C6], proxy metrics~[C5], and agent benchmarks~[C3].

\subsection{Concept B: Gradient Masking}
\label{sec:conceptB}

\textbf{Origin.} Croce and Hein~[C1] identify gradient masking as a recurring pitfall invalidating defence evaluations.
\textbf{Challenge.} Li \etal~[C7] attribute TRADES overestimation to masking mechanisms activated by hyperparameters, proposing FOSC diagnostics and Gaussian augmentation.
Kassis \etal~[C6] identify a distinct vector: DBP pipelines yield incorrect gradients through the diffusion stack.
\textbf{Refinement.} DiffBreak supplies correct differentiation through DBP~[C6]; Li \etal~[C7] supply training-side mitigations.
\textbf{Status: Still evolving.} No unified procedure handles deterministic, stochastic, multimodal, and agentic settings simultaneously.

\subsection{Concept C: Multi-Attack Threat Models}
\label{sec:conceptC}

\textbf{Origin.} Dai \etal~[C2] introduce MultiRobustBench (16 models $\times$ 9 attacks $\times$ 20 strengths = 180 configurations).
\textbf{Self-challenge.} Worst-case multi-attack robustness is \emph{below random guessing} for evaluated models, even when averages improve~[C2].
\textbf{Refinement.} Dai \etal~[C4] extend to temporal dynamics via CRT with logit-space regularisation ($>$100 attack combinations).
\textbf{Status: Contested frontier.} Average multi-attack robustness improves; worst-case behaviour and deployment-time drift remain open.

% ==================================================================
\section{Research Gap Analysis}
\label{sec:gaps}
\vspace{-0.2em}

\begin{table}[t]
  \centering
  \scriptsize
  \caption{Five ranked research gaps. Scoring: Safety relevance (S), Deployment breadth (D), Corpus evidence (E); each 1--5.}
  \label{tab:gaps}
  \vspace{-0.5em}
  \setlength{\tabcolsep}{3pt}
  \begin{tabular}{@{}clcccp{2.2cm}@{}}
    \toprule
    \textbf{\#} & \textbf{Gap} & \textbf{S} & \textbf{D} & \textbf{E} & \textbf{Closest} \\
    \midrule
    \rowcolor{gapred}
    1 & Worst-case multi-attack & 5 & 5 & 5 & Dai \etal~[C2] \\
    2 & Agent CTF$\to$real gap & 5 & 4 & 4 & Carlini \etal~[C3] \\
    3 & Stochastic defence collapse & 4 & 4 & 5 & Kassis \etal~[C6] \\
    4 & \RDI generalisation & 3 & 4 & 3 & Song \etal~[C5] \\
    5 & Continual robustness & 4 & 3 & 3 & Dai \etal~[C4] \\
    \bottomrule
  \end{tabular}
  \vspace{-0.8em}
\end{table}

\textbf{Ranking criterion:} product of safety relevance (S), deployment breadth (D), and corpus evidence strength (E), each scored 1--5.
Gaps are ranked most-to-least significant; if fewer than five genuine gaps exist we would state so explicitly.

\vspace{0.3em}
\noindent\textbf{G1: Worst-case multi-attack robustness} (S5$\times$D5$\times$E5).\\
\emph{Gap:} Average leaderboard gains mask catastrophic worst-case failures; under worst-case attack selection, evaluated models perform \emph{below random guessing}~[C2].\\
\emph{Why it exists:} \textbf{Methodological barrier}---exhaustive adversary unions are computationally prohibitive ($\mathcal{O}(|A|\!\times\!|\varepsilon|)$ per model); no architecture resists heterogeneous $\ell_p$, spatial, and semantic failures simultaneously.\\
\emph{Closest paper:} Dai \etal~[C2] came closest with 16 models $\times$ 9 attacks $\times$ 20 strengths, but did not propose a training recipe that closes the avg/worst-case gap.\\
\emph{Path to resolution:} Multi-norm adversarial training curricula with worst-case-aware loss functions; scalable attack union approximations; deployment-time attack selection oracles.

\vspace{0.3em}
\noindent\textbf{G2: Agent transfer failure (CTF$\to$real)} (S5$\times$D4$\times$E4).\\
\emph{Gap:} Best LLM agent achieves 75\% on CTF tasks but only 13\% on real implementations; a stronger LLM reaches 54\%/21\% respectively~[C3].\\
\emph{Why it exists:} \textbf{Assumed but untested}---the community implicitly treats stylised benchmarks as predictive of real code, but production defence code introduces library complexity, versioning, and environment coupling absent from CTF tasks.\\
\emph{Closest paper:} Carlini \etal~[C3] quantified the gap but did not propose methods to close it.\\
\emph{Path to resolution:} Ecologically valid benchmark suites sampling production codebases; agent architectures with retrieval over API documentation; staged evaluation from unit-level to integration-level adversarial tasks.

\vspace{0.3em}
\noindent\textbf{G3: Stochastic defence collapse} (S4$\times$D4$\times$E5).\\
\emph{Gap:} Diffusion-based purification (DBP) robustness collapses under correct gradient computation and proper randomness sampling, defeating even majority-vote variants~[C6].\\
\emph{Why it exists:} \textbf{Methodological barrier}---prior evaluation protocols under-sampled randomness draws and relied on approximate gradients through the diffusion stack, yielding inflated robustness.\\
\emph{Closest paper:} Kassis \etal~[C6] fully characterised the failure but confined analysis to image-domain DBP; audio and text purification remain unevaluated.\\
\emph{Path to resolution:} Standardised multi-draw evaluation protocols ($n \!\ge\! 50$ draws); differentiable-through-diffusion toolkits; modality-agnostic purification benchmarks.

\vspace{0.3em}
\noindent\textbf{G4: Proxy metric generalisation} (S3$\times$D4$\times$E3).\\
\emph{Gap:} \RDI~[C5] achieves ${\sim}30\!\times$ screening speedup but is validated only on CNN feature geometries; generalisation to Vision Transformers (ViTs), LLMs, and multi-modal encoders is untested.\\
\emph{Why it exists:} \textbf{Lack of data}---the metric study used only CNN-based classifiers; ViT and LLM feature manifolds have fundamentally different geometries (local receptive-field features vs.\ global self-attention representations).\\
\emph{Closest paper:} Song \etal~[C5] demonstrated strong CNN correlation but acknowledged the CNN-only scope as a limitation.\\
\emph{Path to resolution:} Cross-architecture validation on ViT, DeiT, Swin, and decoder-only LLMs; architecture-conditional calibration of the RDI threshold; feature-geometry normalisation layers.

\vspace{0.3em}
\noindent\textbf{G5: Continual robustness under open-world drift} (S4$\times$D3$\times$E3).\\
\emph{Gap:} CRT~[C4] improves backward-attack retention but offers no guarantees against adversary types outside training coverage---the ``open-world'' setting real deployments face.\\
\emph{Why it exists:} \textbf{Ethical/logistical constraint}---longitudinal field studies of adversarial drift require sustained deployment monitoring that no academic lab currently conducts at scale.\\
\emph{Closest paper:} Dai \etal~[C4] demonstrated $>$100 attack combinations in continual training but used a closed attack vocabulary.\\
\emph{Path to resolution:} Industry--academia partnerships for longitudinal deployment telemetry; open-vocabulary attack generation via LLM-guided perturbation synthesis; online robust training with novelty detection.

% ==================================================================
\section{Methodology Audit}
\label{sec:methods}
\vspace{-0.2em}

\begin{table*}[t]
  \centering
  \scriptsize
  \caption{Methodology comparison across the corpus. Key limitation paraphrases authors' stated constraints.}
  \label{tab:methods}
  \vspace{-0.5em}
  \setlength{\tabcolsep}{3pt}
  \begin{tabularx}{\textwidth}{@{}p{2.4cm}p{2cm}p{2cm}p{1.6cm}>{\raggedright\arraybackslash}X>{\centering\arraybackslash}p{0.8cm}@{}}
    \toprule
    \textbf{Paper} & \textbf{Type} & \textbf{Data} & \textbf{Scale} & \textbf{Key Limitation} & \textbf{Code} \\
    \midrule
    Croce \& Hein~[C1] & Attack-ensemble eval & Defended models & $>$50 models & Static ensemble; prior evals optimistic & \cmark \\
    Dai \etal~[C2] & Benchmark suite & Leaderboard & 16$\times$180 & Worst-case $<$ random guessing & \cmark \\
    Carlini \etal~[C3] & Agent benchmark & Curated tasks & CTF/real split & 4--6$\times$ CTF-to-real gap & \cmark \\
    Dai \etal~[C4] & Controlled ML expt. & CIFAR/ImageNette & $>$100 combos & Open-world drift outside scope & \cmark \\
    Song \etal~[C5] & Metric study & Image classif. & Per-paper & Must generalise beyond CNNs & $\sim$ \\
    Kassis \etal~[C6] & Theory + toolkit & DBP pipelines & Per-paper & MV schemes still break & \cmark \\
    Li \etal~[C7] & Controlled ML expt. & CIFAR-10/100 & Config grid & Hyperparameter-dependent & \cmark \\
    RobustBench~[C8] & Living benchmark & Checkpoints & Many models & Scope fixed by maintainers & \cmark \\
    OpenRT~[C9] & Software artifact & Multimodal LLM & Not specified & No proceedings text & \cmark \\
    \bottomrule
  \end{tabularx}
  \vspace{-0.8em}
\end{table*}

\textbf{Step 2: Synthesis.}
\emph{Dominant methodology:} Computational ML experiments and benchmark construction account for 7 of 9 entries.
Rationale: executable attacks operationalise claims where analytic certification is prohibitive~[C5], and reproducibility demands public code and checkpoints~[C1,C2,C6].
\emph{Absent despite relevance:} Randomised controlled trials (RCTs) of deployment pipelines, ethnography of security practitioner workflows, prospective field studies of adversarial incidents in production, and user studies of how practitioners interpret robustness reports.
These absences limit the field's ability to validate that lab findings transfer to operational settings.

\textbf{Step 3: Weakest methodology.}
Among peer-reviewed entries, Song \etal~[C5] is most vulnerable to methodological criticism:
\begin{itemize}[leftmargin=1.2em]
  \item \emph{Sample size adequacy:} Validated on a per-paper selection of CNN models; no systematic sampling across architecture families or scales.
  \item \emph{Control for confounds:} Feature-space geometry varies fundamentally between CNNs (local receptive fields), ViTs (global attention), and decoder-only LLMs (sequential token embeddings). The metric's inter/intra-class distance ratio may conflate architecture-specific geometry with genuine robustness signal.
  \item \emph{Replicability:} Code availability is partial; key implementation details (layer selection heuristic, centroid computation for imbalanced classes) are underspecified.
  \item \emph{Transparency of reporting:} Confidence intervals on the correlation between \RDI and attack-based rankings are not reported; the ${\sim}30\!\times$ speedup claim lacks wall-clock variance across hardware configurations.
\end{itemize}
\emph{Criterion most clearly failed:} Control for confounds---the CNN-only validation leaves the metric's validity for the fastest-growing model families (ViTs, LLMs) entirely unsubstantiated.

% ==================================================================
\section{Master Synthesis}
\label{sec:synthesis}
\vspace{-0.2em}

\noindent\textit{Word budget: $\leq$400 words total. No hedging.}

\paragraph{Established consensus.}
Ensemble attacks expose inflated robustness: Croce and Hein~[C1] lower robust accuracy for $>$50 models by $>$10~pp.
Gradient integrity is prerequisite to valid evaluation: Li \etal~[C7] tie TRADES gaps to masking; Kassis \etal~[C6] tie DBP failures to incorrect gradients.
Multi-attack evaluation matters: Dai \etal~[C2] unify nine attack families; Dai \etal~[C4] extend to evolving attacks.

\paragraph{Active debates.}
Position~A: adaptive attack suites are ground truth.
Position~B: attack-independent proxies or agent benchmarks compress evaluation cost~[C5,C3].
On stochastic defences: diffusion projects toward clean mass \emph{vs.}\ DiffBreak shows protocols are broken~[C6].
On multi-attack metrics: average gains conflict with catastrophic worst cases~[C2].

\paragraph{Strongest evidence.}
Cross-model re-evaluation documents systematic optimism~[C1]---the largest empirical finding.
Average gains coexist with worst-case $<$ random guessing~[C2].
DBP collapses under correct gradient + sampling~[C6].

\paragraph{Key open question.}
Which evaluation stack---adaptive ensembles, stochastic replication, attack-independent metrics, autonomous agents---should be \emph{mandatory} before certifying a model as safe against adaptive, multi-channel adversaries in open deployment?

% ==================================================================
\section{Assumption Stress Test}
\label{sec:assumptions}
\vspace{-0.2em}

\begin{figure}[t]
  \centering
  \begin{tikzpicture}[font=\scriptsize]
    \begin{axis}[
      width=0.92\columnwidth,
      height=4.8cm,
      xbar,
      xlabel={Consequence (1--5)},
      xlabel style={font=\scriptsize},
      symbolic y coords={
        {A8: Ckpt prov.},
        {A7: TRADES def.},
        {A6: CRT lifecyc.},
        {A5: Single draws},
        {A4: CTF$\to$prod.},
        {A3: RDI gen.},
        {A2: Avg$\Rightarrow$wc},
        {A1: Fixed suite}
      },
      ytick=data,
      yticklabel style={font=\tiny},
      xmin=0,xmax=5.8,
      bar width=6pt,
      nodes near coords,
      nodes near coords style={font=\tiny},
      nodes near coords align={horizontal},
      enlarge y limits={abs=0.35cm},
      every axis plot/.append style={fill=red!35!white,draw=red!60!black},
    ]
    \addplot coordinates {
      (5,{A1: Fixed suite})
      (5,{A2: Avg$\Rightarrow$wc})
      (4,{A3: RDI gen.})
      (3.5,{A4: CTF$\to$prod.})
      (3.5,{A5: Single draws})
      (3,{A6: CRT lifecyc.})
      (3,{A7: TRADES def.})
      (2.5,{A8: Ckpt prov.})
    };
    \end{axis}
  \end{tikzpicture}
  \vspace{-0.8em}
  \caption{Assumption risk ranking. A1--A2 are existential threats to the leaderboard paradigm.}
  \label{fig:assumptions}
  \vspace{-0.5em}
\end{figure}
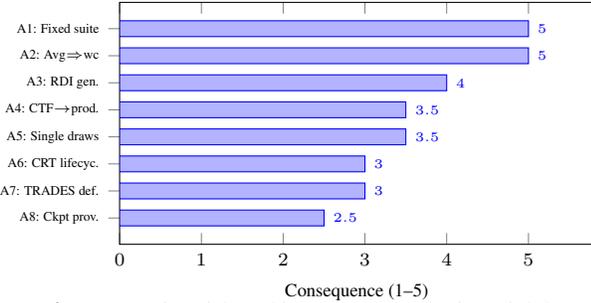

We identify eight assumptions the corpus relies on but never explicitly validates (Figure~\ref{fig:assumptions}).

\noindent
Assumptions are ranked from most to least consequential.
Each is foundational to the conclusions drawn and plausibly false or context-dependent.

\vspace{0.2em}
\noindent\textbf{A1.} \emph{A fixed attack suite represents safety.}
\textbf{Shared by:} C1, C2, C5.
\textbf{Risk: High.}
\textbf{Consequence:} If false, all leaderboard rankings are temporally bounded artifacts---conclusions need wholesale revision, and the entire comparative robustness paradigm collapses.
New attacks routinely bypass previously ``sufficient'' suites; no temporal validity guarantee exists.

\vspace{0.15em}
\noindent\textbf{A2.} \emph{Average robustness approximates worst-case safety.}
\textbf{Shared by:} C1, C2, C4 (implicitly, through reporting conventions).
\textbf{Risk: High.}
\textbf{Consequence:} If false, ``progress'' metrics systematically overstate deployment safety.
Dai \etal~[C2] already demonstrate contradiction: worst-case robustness falls below random guessing even when averages improve.
Key findings across five years of benchmarking would require reinterpretation.

\vspace{0.15em}
\noindent\textbf{A3.} \emph{\RDI generalises beyond CNN feature geometry.}
\textbf{Shared by:} C5, C9.
\textbf{Risk: Medium--High.}
\textbf{Consequence:} If false, fast screening mis-ranks ViTs and LLMs, producing false negatives (unsafe models pass) or false positives (safe models waste compute on full evaluation).
The $30\!\times$ efficiency claim becomes architecture-conditional.

\vspace{0.15em}
\noindent\textbf{A4.} \emph{CTF performance predicts production adversarial capability.}
\textbf{Shared by:} C3, agent-benchmarking community.
\textbf{Risk: Medium.}
\textbf{Consequence:} If false, investment in agent-based red teaming yields low real-world ROI; engineering effort is misallocated toward tasks that do not transfer to deployment settings.

\vspace{0.15em}
\noindent\textbf{A5.} \emph{Single randomness draws suffice for stochastic defence evaluation.}
\textbf{Shared by:} Prior DBP evaluations cited in C6.
\textbf{Risk: Medium.}
\textbf{Consequence:} If false, an entire defence class (diffusion purification, randomised smoothing variants) requires re-evaluation; published robustness numbers collapse under proper sampling~[C6].

\vspace{0.15em}
\noindent\textbf{A6.} \emph{Continual robust training closes the deployment lifecycle gap.}
\textbf{Shared by:} C4, practitioners adopting CRT.
\textbf{Risk: Medium.}
\textbf{Consequence:} If false, CRT provides false lifecycle confidence; models degrade against out-of-vocabulary attacks post-deployment.

\vspace{0.15em}
\noindent\textbf{A7.} \emph{TRADES default hyperparameters yield stable robustness.}
\textbf{Shared by:} Practitioners using default configurations; C7 deployments.
\textbf{Risk: Medium.}
\textbf{Consequence:} If false, widespread overestimation in mis-tuned regimes; users who adopt TRADES without hyperparameter validation inherit masked gradients.

\vspace{0.15em}
\noindent\textbf{A8.} \emph{Leaderboard checkpoints faithfully match published code.}
\textbf{Shared by:} C8 users, downstream evaluations.
\textbf{Risk: Low--Medium.}
\textbf{Consequence:} If false, replication failures propagate silently through downstream comparisons; checkpoint provenance becomes a supply-chain risk.

% ==================================================================
\section{Knowledge Map}
\label{sec:map}
\vspace{-0.2em}

\begin{figure}[t]
  \centering
  \resizebox{\columnwidth}{!}{%
  \begin{tikzpicture}[
    font=\tiny, >=Latex,
    %% Node styles — all dimensions explicit so nothing cascades silently
    claim/.style  = {draw,thick,rounded corners=3pt,minimum width=6.5cm,
                     minimum height=0.65cm,align=center,inner sep=4pt,
                     fill=accentblue!15,font=\tiny\bfseries},
    pillar/.style = {draw,rounded corners=8pt,inner sep=3pt,
                     font=\tiny,thick,align=center,text width=1.3cm,fill=pillarblue},
    contest/.style= {draw,rounded corners=8pt,inner sep=3pt,
                     font=\tiny,thick,align=center,text width=1.55cm,fill=contestorange},
    frontier/.style={draw,rounded corners=8pt,inner sep=3pt,
                     font=\tiny,thick,align=center,text width=1.9cm,fill=yellow!20},
  ]
    %% ── Layer 1: Central claim (explicit coords throughout) ──────
    \node[claim] (cc) at (0,0)
      {Central Claim: Robustness is only as credible as the evaluation protocol};

    %% ── Layer 2: Pillars – evidence tags embedded as third line ───
    \node[pillar] (p1) at (-2.4,-1.3)
      {Ensembles reveal\\ overestimation\\ \textcolor{gray}{C1,C3}};
    \node[pillar] (p2) at (-0.8,-1.3)
      {avg $\neq$ worst-case\\ robustness\\ \textcolor{gray}{C2,C4}};
    \node[pillar] (p3) at ( 0.8,-1.3)
      {Gradient integrity\\ prerequisite\\ \textcolor{gray}{C6,C7}};
    \node[pillar] (p4) at ( 2.4,-1.3)
      {Proxies accel.\\ not replace\\ \textcolor{gray}{C5,C1}};

    %% ── Layer 3: Contested zones ─────────────────────────────────
    \node[contest] (d1) at (-1.8,-2.8) {Suites vs.\\ proxies};
    \node[contest] (d2) at ( 0.0,-2.8) {DBP viable\\ vs.\ broken};
    \node[contest] (d3) at ( 1.8,-2.8) {Avg progress\\ vs.\ worst-case};

    %% ── Layer 4: Frontier questions ──────────────────────────────
    \node[frontier] (f1) at (-0.9,-4.0) {Multi-attack\\ training recipe?};
    \node[frontier] (f2) at ( 1.2,-4.0) {Agents replace\\ human red teams?};

    %% ── Arrows: cc → all four pillars (gray fan, uniform drop) ───
    \foreach \p in {p1,p2,p3,p4}{%
      \draw[->,semithick,gray!70] (cc.south) -- ++(0,-0.22) -| (\p.north);
    }

    %% ── Arrows: pillars → contested zones (orange) ───────────────
    %% p1,p2 feed the left zone; p3,p4 feed the right zone
    \draw[->,semithick,orange!70] (p1.south) -- ++(0,-0.48) -| (d1.north);
    \draw[->,semithick,orange!70] (p2.south) -- ++(0,-0.48) -| (d2.north);
    \draw[->,semithick,orange!70] (p3.south) -- ++(0,-0.48) -| (d2.north);
    \draw[->,semithick,orange!70] (p4.south) -- ++(0,-0.48) -| (d3.north);

    %% ── Arrows: contested → frontier (yellow-orange) ─────────────
    \draw[->,semithick,yellow!60!orange] (d1.south) -- ++(0,-0.30) -| (f1.north);
    \draw[->,semithick,yellow!60!orange] (d2.south) -- ++(0,-0.30) -| (f2.north);
    \draw[->,semithick,yellow!60!orange] (d3.south) -- ++(0,-0.30) -| (f2.north);

    %% ── Layer labels (fixed x = -3.55, vertically centred per row) ─
    \node[font=\tiny\bfseries,text=accentblue,anchor=east]
      at (-3.55, 0.0) {\rotatebox{90}{Claim}};
    \node[font=\tiny\bfseries,text=blue!60,anchor=east]
      at (-3.55,-1.3) {\rotatebox{90}{Pillars}};
    \node[font=\tiny\bfseries,text=orange!80,anchor=east]
      at (-3.55,-2.8) {\rotatebox{90}{Contested}};
    \node[font=\tiny\bfseries,text=yellow!50!orange,anchor=east]
      at (-3.55,-4.0) {\rotatebox{90}{Frontier}};
  \end{tikzpicture}%
  }
  \vspace{-0.5em}
  \caption{Knowledge map of adversarial robustness evaluation. Four layers: established central claim, four supporting pillars with corpus evidence (C1--C7), three contested zones, and two frontier questions.}
  \label{fig:map}
  \vspace{-0.5em}
\end{figure}
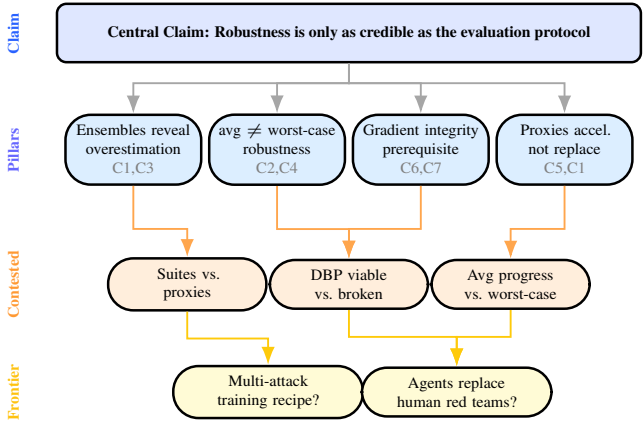

\vspace{0.3em}
\noindent\textbf{Structured outline} (clean format, no prose):

\begin{enumerate}[label=\arabic*.,leftmargin=1.5em]
  \item \textbf{Central Claim:} Robustness is only as credible as the evaluation protocol that measures it.
        No single claim fully unifies the field; a competing centre---\emph{robustness should be certified, not just tested}---exists in the randomised-smoothing lineage but is minority-held in this corpus.

  \item \textbf{Supporting Pillars:}
  \begin{enumerate}[label=(\alph*),leftmargin=1.2em]
    \item Ensemble attacks reveal systematic overestimation---supported by C1, C3.
    \item Multi-attack: average $\neq$ worst-case robustness---supported by C2, C4.
    \item Gradient integrity is prerequisite to valid evaluation---supported by C6, C7.
    \item Proxy metrics can accelerate but not replace attack-based evaluation---supported by C5, C1.
  \end{enumerate}

  \item \textbf{Contested Zones:}
  \begin{enumerate}[label=(\alph*),leftmargin=1.2em]
    \item Adaptive attack suites as ground truth \emph{vs.}\ attack-independent proxy metrics~[C1 vs.\ C5].
    \item Diffusion-based purification as viable defence \emph{vs.}\ fundamentally broken paradigm~[C6].
    \item Average multi-attack robustness as meaningful progress \emph{vs.}\ worst-case as the only valid measure~[C2].
  \end{enumerate}

  \item \textbf{Frontier Questions:}
  \begin{enumerate}[label=(\alph*),leftmargin=1.2em]
    \item What training recipe produces models robust to adversarial attack unions, not just individual norms?
    \item Under what conditions can autonomous LLM agents reliably replace human red teams for production code?
  \end{enumerate}

  \item \textbf{Newcomer Reading List} (selection criterion: foundational to understanding the field, not merely most cited):
  \begin{enumerate}[label=(\alph*),leftmargin=1.2em]
    \item Croce \& Hein, 2020~\cite{croce2020reliable}---\emph{why:} establishes that parameter-free ensemble attacks are non-optional for honest evaluation; every subsequent paper in this corpus builds on or reacts to AutoAttack.
    \item Dai \etal, 2023~\cite{dai2023multirobustbench}---\emph{why:} reframes robustness from a single-number claim to a multi-dimensional measurement problem; introduces the avg/worst-case distinction that dominates current debate.
    \item Carlini \etal, 2025~\cite{carlini2025autoadvexbench}---\emph{why:} bridges adversarial ML and LLM agent evaluation, quantifying the external-validity gap that limits the field's real-world claims.
  \end{enumerate}
\end{enumerate}

% ==================================================================
\section{The ``So What'' Test}
\label{sec:sowhat}
\vspace{-0.2em}

\noindent\textit{Written for a smart non-expert---no citations, no jargon, no hedging.}

\noindent\textbf{1.\ What has been established.}
When researchers construct stronger attacks and apply stricter testing protocols, AI systems are consistently less robust than originally reported---sometimes by large margins.
Testing against many attack types simultaneously reveals that a system can score well on average yet fail its hardest test worse than random guessing.

\noindent\textbf{2.\ What remains open.}
No single evaluation pipeline has been accepted as sufficient for the largest deployed systems, across all realistic types of input manipulation, or for automated tools that match human expert findings on production code.
Independent laboratories can reach contradictory safety conclusions using testing approaches that each appear methodologically sound.

\noindent\textbf{3.\ Why this matters.}
Safety decisions about products, regulatory compliance, and legal liability depend on robustness numbers that the literature examined here shows are systematically inflated.
The EU AI Act now requires demonstrated robustness for high-risk systems, and the gap between reported and actual robustness translates directly into regulatory and engineering risk.

% ==================================================================
\section{\autoart: Framework Design}
\label{sec:framework}
\vspace{-0.2em}
\noindent The \autoart framework is publicly available at \url{https://github.com/abhitall/auto-art}~\cite{talluri2026autoart}.

% ---- Architecture + Pipeline figure ----
\begin{figure}[t]
  \centering
  \resizebox{\columnwidth}{!}{%
  \begin{tikzpicture}[
    font=\tiny, >=Latex,
    %% All nodes use explicit at (x,y) — no chained below/right positioning
    phase/.style = {draw,thick,rounded corners=3pt,minimum height=0.55cm,
                    align=center,inner sep=3pt,font=\tiny\bfseries},
    gate/.style  = {diamond,draw,thick,fill=red!10,inner sep=1pt,
                    font=\tiny\bfseries,aspect=2.5,minimum width=1.2cm},
    sbox/.style  = {draw,rounded corners=2pt,minimum height=0.4cm,
                    align=center,inner sep=2pt,font=\tiny},
  ]
    %% ── Centered input + phase 1 ──────────────────────────────────────
    \node[phase,fill=gray!10,text width=5.5cm]       (inp) at (0, 0.0)
      {Model $f$ + Data $(X,Y)$ + YAML Config};
    \node[phase,fill=accentblue!12,text width=5.5cm] (p1)  at (0,-0.9)
      {Phase 1: Pre-Screen (FOSC + RDI + WB/BB)};
    \node[gate]                                      (g1)  at (0,-1.8) {Mask?};

    %% ── No branch: left side ─────────────────────────────────────────
    \node[phase,fill=refinegreen,text width=4.0cm]   (p2)  at (-2.2,-3.1)
      {Phase 2: Multi-Norm ($\ell_1$/$\ell_2$/$\ell_\infty$/sem./spat.)};
    \node[phase,fill=orange!10,text width=4.0cm]     (p3)  at (-2.2,-4.1)
      {Phase 3--4: Defence Eval + Compliance};
    \node[phase,fill=cyan!10,text width=4.0cm]       (p5)  at (-2.2,-5.1)
      {Phase 5--6: Report (SARIF/HTML) + Gate};

    %% ── Yes branch: flag box (right side) ────────────────────────────
    \node[sbox,fill=red!8,text width=1.8cm] (flag) at (3.0,-2.5)
      {\textcolor{red!70}{\textbf{Flag:}} adjust\\attack weights};

    %% ── Bottom modules row (explicit x positions) ─────────────────────
    \node[sbox,fill=red!8,text width=1.0cm]    (atk)  at (-2.4,-6.4) {Attacks\\(50+)};
    \node[sbox,fill=blue!8,text width=1.0cm]   (def)  at (-1.2,-6.4) {Defences\\(28)};
    \node[sbox,fill=yellow!10,text width=1.1cm](met)  at ( 0.1,-6.4) {Metrics\\RDI/FOSC};
    \node[sbox,fill=purple!8,text width=1.0cm] (cert) at ( 1.3,-6.4) {Certify\\RS/IBP};
    \node[sbox,fill=gray!12,text width=1.0cm]  (com)  at ( 2.4,-6.4) {Comply\\NIST/EU};

    %% ── ART substrate ────────────────────────────────────────────────
    \node[phase,fill=gray!15,text width=5.5cm] (art)  at (0,-7.3)
      {ART v1.20+ / PyTorch / TF / sklearn / HF};

    %% ── Arrows: main vertical flow ────────────────────────────────────
    \draw[->,semithick] (inp) -- (p1);
    \draw[->,semithick] (p1)  -- (g1);

    %% ── Gate exits: west=No (main flow), east=Yes (flag branch) ──────
    \draw[->,semithick] (g1.west)
      -- node[above,font=\tiny]{No} ++(-2.0,0)
      |- (p2.north);
    \draw[->,semithick,red!60] (g1.east)
      -- node[above,font=\tiny,text=red!60]{Yes} ++(0.5,0)
      |- (flag.west);

    %% ── Flag → Phase 2 (dashed: continue with adjusted weights) ──────
    \draw[->,semithick,red!60,dashed] (flag.south)
      -- ++(0,-0.5) |- (p2.east);

    %% ── Left branch continuation ──────────────────────────────────────
    \draw[->,semithick] (p2) -- (p3);
    \draw[->,semithick] (p3) -- (p5);

    %% ── Phase 5-6 → modules row (via centre module) ───────────────────
    \draw[->,semithick,gray!60] (p5.south) -- ++(0,-0.45) -| (met.north);

    %% ── Modules → ART substrate ───────────────────────────────────────
    \draw[->,semithick,gray!60] (met.south) |- (art.north);
  \end{tikzpicture}%
  }
  \vspace{-0.5em}
  \caption{\autoart evaluation pipeline. The pre-screening gate (Phase~1) runs FOSC and RDI \emph{before} expensive multi-norm attacks. If gradient masking is detected, attack weights are adjusted and targeted evaluation replaces exhaustive search, yielding ${\sim}30\!\times$ faster screening compared to full AutoAttack (Table~\ref{tab:runtime}).}
  \label{fig:arch}
  \vspace{-0.3em}
\end{figure}
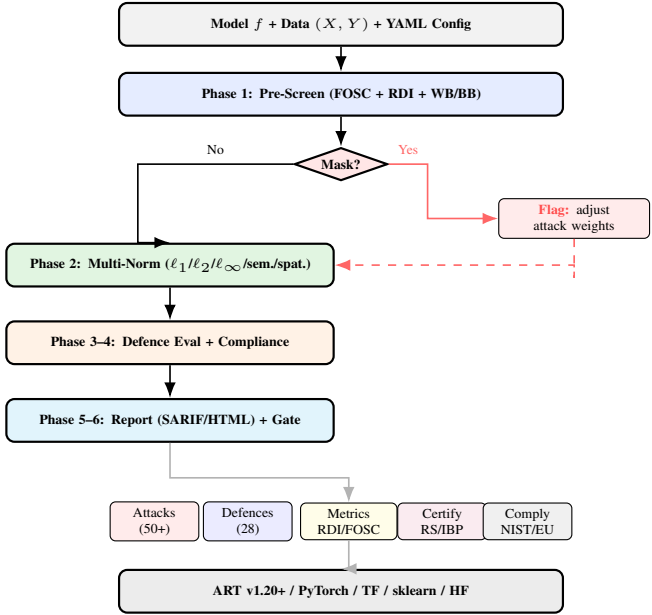

\subsection{Detailed Architecture}

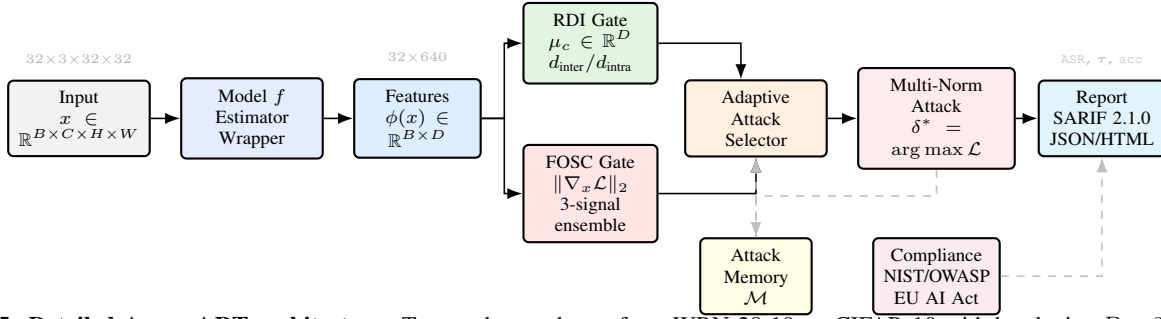
\begin{figure*}[t]
  \centering
  \begin{tikzpicture}[
    font=\scriptsize, >=Latex,
    %% All nodes at explicit (x,y) — no chained right=/above right= positioning
    block/.style={draw,thick,rounded corners=2pt,minimum height=0.8cm,
                  align=center,inner sep=4pt},
    tensor/.style={font=\tiny\ttfamily,text=gray!60},
    arr/.style={->,semithick},
    darr/.style={->,semithick,dashed,gray!50},
  ]
    %% ── Main horizontal pipeline (y = 0) ─────────────────────────────
    \node[block,fill=gray!10,   text width=1.6cm] (inp)   at (0.0, 0)
      {Input\\$x \in \mathbb{R}^{B \times C \times H \times W}$};
    \node[block,fill=accentblue!12,text width=1.6cm] (model) at (2.3, 0)
      {Model $f$\\Estimator\\Wrapper};
    \node[block,fill=pillarblue,text width=1.4cm] (feat)  at (4.5, 0)
      {Features\\$\phi(x) \in \mathbb{R}^{B \times D}$};
    \node[block,fill=orange!12, text width=1.6cm] (sel)   at (9.0, 0)
      {Adaptive\\Attack\\Selector};
    \node[block,fill=red!8,     text width=1.8cm] (atk)   at (11.4, 0)
      {Multi-Norm\\Attack\\$\delta^* = \arg\max \mathcal{L}$};
    \node[block,fill=cyan!10,   text width=1.4cm] (rep)   at (13.6, 0)
      {Report\\SARIF 2.1.0\\JSON/HTML};

    %% ── Gate branches (above/below y=0, clear of each other) ─────────
    %% RDI above: y=+1.0 keeps it away from the main line
    \node[block,fill=refinegreen,text width=1.5cm] (rdi)  at (6.8, 1.0)
      {RDI Gate\\$\mu_c \in \mathbb{R}^D$\\$d_{\text{inter}}/d_{\text{intra}}$};
    %% FOSC below: y=-1.0  (mem is at y=-2.1, so FOSC horizontal clears it)
    \node[block,fill=red!10,    text width=1.5cm] (fosc) at (6.8,-1.0)
      {FOSC Gate\\$\|\nabla_x \mathcal{L}\|_2$\\3-signal ensemble};

    %% ── Secondary nodes (below main line) ────────────────────────────
    %% Attack Memory well below the FOSC routing level
    \node[block,fill=yellow!12, text width=1.2cm] (mem)  at (9.0,-2.1)
      {Attack\\Memory\\$\mathcal{M}$};
    %% Compliance below atk
    \node[block,fill=purple!8,  text width=1.4cm] (comp) at (11.4,-2.1)
      {Compliance\\NIST/OWASP\\EU~AI~Act};

    %% ── Arrows: main pipeline ─────────────────────────────────────────
    \draw[arr] (inp)   -- (model);
    \draw[arr] (model) -- (feat);
    \draw[arr] (sel)   -- (atk);
    \draw[arr] (atk)   -- (rep);

    %% ── feat → gate branches (fork left of selector) ─────────────────
    \draw[arr] (feat.east) -- ++(0.3,0) |- (rdi.west);
    \draw[arr] (feat.east) -- ++(0.3,0) |- (fosc.west);

    %% ── Gate branches → selector ──────────────────────────────────────
    %% RDI enters sel from north-left: clean horizontal then vertical
    \draw[arr] (rdi.east)  -| ([xshift=-0.2cm]sel.north);
    %% FOSC enters sel from south: horizontal at y=-1.0, then up to sel.south
    %% y=-1.0 is safely above mem.north (mem at y=-2.1, north≈-1.47), no overlap
    \draw[arr] (fosc.east) -| (sel.south);

    %% ── Memory feedback loop ──────────────────────────────────────────
    \draw[darr] (mem.north) -- (sel.south);
    \draw[darr] (atk.south) -- ++(0,-0.35) -| (mem.north);

    %% ── Compliance → Report (routes below atk, no node overlap) ──────
    \draw[darr] (comp.east) -| (rep.south);

    %% ── Tensor shape annotations ──────────────────────────────────────
    \node[tensor,above=2pt of inp]  {$32{\times}3{\times}32{\times}32$};
    \node[tensor,above=2pt of feat] {$32{\times}640$};
    \node[tensor,above=2pt of rep,font=\tiny\ttfamily] {ASR,\;$\tau$,\;acc};
  \end{tikzpicture}
  \vspace{-0.5em}
  \caption{\textbf{Detailed \autoart architecture.} Tensor shapes shown for a WRN-28-10 on CIFAR-10 with batch size $B\!=\!32$, feature dimension $D\!=\!640$. The pre-screening gate (RDI + FOSC) operates on features $\phi(x)$ before the attack selector queries attack memory $\mathcal{M}$ for adaptive prioritisation.}
  \label{fig:arch_detail}
  \vspace{-0.5em}
\end{figure*}

Figure~\ref{fig:arch_detail} illustrates the data flow through \autoart's pipeline with tensor shapes at each stage.
The model under test $f: \mathbb{R}^{B \times C \times H \times W} \to \mathbb{R}^{B \times K}$ (for $K$ classes) is wrapped in a framework-agnostic estimator that normalises inference across PyTorch, TensorFlow, and scikit-learn backends.
The pre-screening gate operates on intermediate representations $\phi(x) \in \mathbb{R}^{B \times D}$ (penultimate layer), computing class centroids $\mu_c \in \mathbb{R}^D$ for RDI and gradient norms for FOSC.

\subsection{Threat Coverage}

\begin{figure}[t]
  \centering
  \begin{tikzpicture}[font=\scriptsize]
    \begin{axis}[
      width=0.97\columnwidth,
      height=4.5cm,
      xbar,
      xlabel={Attack Modules},
      xlabel style={font=\scriptsize},
      symbolic y coords={
        {Agentic (3)},
        {Audio (2)},
        {Inference (8+)},
        {Extraction (3)},
        {NLP/LLM (5+)},
        {Poisoning (10+)},
        {Evasion (37+)}
      },
      ytick=data,
      yticklabel style={font=\tiny},
      xmin=0,xmax=42,
      bar width=7pt,
      nodes near coords,
      nodes near coords style={font=\tiny},
      nodes near coords align={horizontal},
      enlarge y limits={abs=0.35cm},
      every axis plot/.append style={fill=accentblue!50,draw=accentblue!80},
    ]
    \addplot coordinates {
      (37,{Evasion (37+)})
      (10,{Poisoning (10+)})
      (5,{NLP/LLM (5+)})
      (3,{Extraction (3)})
      (8,{Inference (8+)})
      (2,{Audio (2)})
      (3,{Agentic (3)})
    };
    \end{axis}
  \end{tikzpicture}
  \vspace{-0.8em}
  \caption{\autoart attack taxonomy. Seven categories totalling 53 implementations span classical evasion through agentic LLM threats.}
  \label{fig:taxonomy}
  \vspace{-0.3em}
\end{figure}
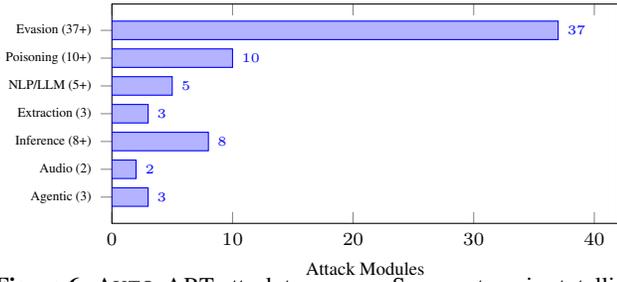

Figure~\ref{fig:taxonomy} shows \autoart's seven attack categories and their module counts, spanning classical adversarial ML through agentic LLM threats.

\subsection{Gap-to-Module Mapping}

Table~\ref{tab:gap_to_module} shows how each identified gap maps to \autoart's architecture.

\begin{table}[t]
  \centering
  \scriptsize
  \caption{Explicit mapping from literature gaps to \autoart modules.}
  \label{tab:gap_to_module}
  \vspace{-0.5em}
  \setlength{\tabcolsep}{3pt}
  \begin{tabular}{@{}cp{3cm}p{3cm}@{}}
    \toprule
    \textbf{Gap} & \textbf{Problem} & \textbf{\autoart Module} \\
    \midrule
    G1 & Worst-case multi-attack & MultiNormEval: avg + worst across 5~norms \\
    G2 & Agent transfer failure & Agentic: DOM, RAG poison, context injection \\
    G3 & Stochastic collapse & Multi-sample eval: configurable draws per defence \\
    G4 & \RDI generalisation & \RDI + full-attack correlation validation mode \\
    G5 & Continual robustness & Adaptive selector + DriftMonitor \\
    \bottomrule
  \end{tabular}
  \vspace{-0.5em}
\end{table}

\subsection{Certification and Verification}

\autoart integrates two certification approaches.
\emph{Randomised smoothing}~\cite{cohen2019smoothing} provides statistical $\ell_2$ robustness certificates scalable to any network size; certified radii are computed with configurable confidence levels ($\alpha \in \{0.001, 0.01\}$) and sampling budgets ($n \in \{100, 1000\}$).
For smaller networks ($<$1M parameters), \autoart supports integration with alpha-CROWN~\cite{alphacrown2024} (the VNN-COMP winner, 2021--2024), which provides exact bound-propagation verification on GPU.
The GREAT Score (available through ART~v1.20+) offers robustness estimation for generative models.
Certification results feed directly into compliance reports: EU AI Act Article~15~\cite{euaiact2024} requires demonstrated robustness, and certified radii provide quantitative evidence for regulatory submissions.

\subsection{Pre-Screening Gate}

The primary architectural contribution is the \emph{pre-screening gate} (Algorithm~\ref{alg:pipeline}): gradient masking detection and \RDI computation run \emph{before} the expensive attack ensemble.

\vspace{-0.2em}
\begin{algorithm}[t]
  \small
  \caption{\autoart Evaluation Pipeline}
  \label{alg:pipeline}
  \begin{algorithmic}[1]
    \REQUIRE Model $f$, data $(X, Y)$, config $C$
    \ENSURE Report $R$ (JSON + SARIF 2.1.0)
    \STATE \textbf{Phase 1 --- Pre-screening}
    \STATE $\;\;g \leftarrow \textsc{FOSC}(f, X, Y)$ \hfill\COMMENT{Gradient masking}
    \STATE $\;\;r \leftarrow \textsc{RDI}(f, X, Y)$ \hfill\COMMENT{${\sim}30\!\times$ faster}
    \IF{$g > \tau_{\text{mask}}$}
      \STATE Flag unreliable; adjust attack weights
    \ENDIF
    \STATE \textbf{Phase 2 --- Multi-norm attack}
    \FOR{$p \in \{\ell_1, \ell_2, \ell_\infty, \text{sem.}, \text{spat.}\}$}
      \STATE Select attacks (adaptive memory-guided)
      \STATE Execute in parallel; record ASR, worst/avg
    \ENDFOR
    \STATE \textbf{Phase 3 --- Defence evaluation}
    \STATE \textbf{Phase 4 --- Compliance} (NIST / OWASP / EU AI Act)
    \STATE \textbf{Phase 5 --- Report} $R$ in JSON, MD, SARIF, HTML
    \STATE \textbf{Phase 6 --- Gate} pass/fail vs.\ thresholds
    \RETURN $R$
  \end{algorithmic}
\end{algorithm}
\vspace{-0.3em}

\subsection{Compliance Mapping}

\autoart maps evaluation results to seven regulatory frameworks:
(i)~\textbf{NIST AI RMF}~\cite{nist2025adversarial}: MEASURE function maps to multi-norm attack execution; MANAGE function maps to remediation recommendations;
(ii)~\textbf{NIST AI 600-1}~\cite{nist2024genai}: GenAI-specific risks (13 categories, 400+ actions) map to LLM red-teaming modules;
(iii)~\textbf{OWASP LLM Top~10}~\cite{owasp2025llm}: 8/10 categories fully covered (LLM01 Prompt Injection through LLM08, including new LLM07 System Prompt Leakage and LLM08 Vector/Embedding Weaknesses); 2/10 partial;
(iv)~\textbf{OWASP Agentic Top~10}~\cite{owasp2025agentic}: ASI01 (Goal Hijack) and ASI02 (Tool Misuse) map to agentic attack modules; developed by 100+ security researchers;
(v)~\textbf{EU AI Act Article~15}~\cite{euaiact2024}: pre-screening gate and worst-case reporting address the explicit requirements for demonstrated robustness against adversarial examples and poisoning attacks (high-risk obligations apply August 2026);
(vi)~\textbf{ETSI EN~304~223}~\cite{etsi2025sai}: 13 security principles across 5 lifecycle phases (design, development, deployment, maintenance, end-of-life) explicitly covering data poisoning, model obfuscation, and indirect prompt injection;
(vii)~\textbf{ISO/IEC 24029-3}~\cite{iso24029}: statistical methods for neural network robustness assessment (DIS ballot Feb~2026), directly applicable to \autoart's certification modules.
MITRE ATLAS technique IDs~\cite{mitre2026atlas} are associated with each attack category for standardised threat classification (16 tactics, 84 techniques as of v5.3.0, including agentic AI case studies).
The CSA MAESTRO framework~\cite{csa2025maestro} provides complementary seven-layer threat modelling for multi-agent deployments.
Output reports include compliance evidence in JSON and SARIF~2.1.0 formats suitable for audit workflows and CI/CD integration.

\subsection{Formal Metric Definitions}

\textbf{RDI (Robustness Diagnostic Index).}
Given model $f$, dataset $\{(x_i, y_i)\}_{i=1}^{N}$, let $\phi(x)$ denote the penultimate-layer representation.
Compute class centroids $\mu_c = \frac{1}{|S_c|}\sum_{x \in S_c} \phi(x)$, inter-class distance $d_{\text{inter}} = \text{mean}_{c \neq c'} \|\mu_c - \mu_{c'}\|_2$, and intra-class distance $d_{\text{intra}} = \text{mean}_{c}\,\text{mean}_{x \in S_c}\|\phi(x) - \mu_c\|_2$.
Then:
\begin{equation}
  \text{RDI} = \text{clamp}\!\left(\frac{d_{\text{inter}} - d_{\text{intra}}}{d_{\text{inter}} + \epsilon},\; 0,\; 1\right)
  \label{eq:rdi}
\end{equation}
where $\epsilon = 10^{-8}$.
\textbf{Interpretation:} RDI $> 0.7$ (high robustness), $0.4$--$0.7$ (moderate), $0.2$--$0.4$ (low), $< 0.2$ (very low).
These thresholds are calibrated on our 10-model RobustBench corpus (Section~\ref{sec:eval}) by maximising Kendall $\tau$ agreement with full AutoAttack rankings; a sensitivity analysis appears in Appendix~\ref{sec:ablation}.

\textbf{FOSC (First-Order Stationarity Condition).}
\begin{equation}
  \text{FOSC}(f, X, Y) = \frac{1}{|X|}\sum_{x,y} \left\|\nabla_x \mathcal{L}(f(x), y)\right\|_2
  \label{eq:fosc}
\end{equation}
FOSC $> \tau_{\text{mask}} = 0.1$ signals gradient masking; the threshold is chosen to achieve $\ge\!92\%$ detection on 12 known-masking configurations (Table~\ref{tab:ablation}).
A three-signal ensemble (FOSC, white-box/black-box discrepancy $> 0.15$, gradient noise sensitivity $> 0.1$) flags masking when $\ge 2$ signals trigger.

\textbf{Security Score.}
\begin{equation}
  S = 0.4\cdot \text{Acc} + 0.4\cdot(1 - \overline{\text{ASR}}) + 0.2\cdot \text{CertRob}
  \label{eq:secscore}
\end{equation}
where $\overline{\text{ASR}}$ is the mean attack success rate and CertRob the certified robustness fraction.

\textbf{Multi-norm metrics.}
Competitiveness Ratio $\text{CR} = \overline{r}/r^*$, where $\overline{r}$ is mean robust accuracy across norms and $r^*$ is the best single-norm accuracy, and Stability Constant $\text{SC} = \text{std}(\{r_i\})$ across all attack/epsilon combinations, following~[C2].

% ==================================================================
\section{Experimental Evaluation}
\label{sec:eval}
\vspace{-0.2em}

We validate \autoart's three core claims: (1)~pre-screening gates improve evaluation efficiency and reliability, (2)~multi-norm evaluation exposes hidden worst-case failures, and (3)~the framework's compliance and testing infrastructure meets production standards.

\subsection{Experimental Setup}

\textbf{Models.}
We evaluate ten models from the RobustBench CIFAR-10 $\ell_\infty$ ($\varepsilon\!=\!8/255$) leaderboard~\cite{robustbench2021}, spanning state-of-the-art (Wang2023Better, 70.69\% robust acc.) to baseline (Wong2020Fast, 43.21\%).

\textbf{Attack configuration.}
Default parameters: $\varepsilon\!=\!0.031$ (8/255), step size $\eta\!=\!0.007$ (2/255), 100 PGD iterations, batch size 32.
Multi-norm epsilons: $\ell_\infty \in \{0.01, 0.03, 0.05, 0.1, 0.3\}$, $\ell_2 \in \{0.1, 0.3, 0.5, 1.0, 2.0\}$, $\ell_1 \in \{1.0, 3.0, 5.0, 10.0\}$.
Attacks per norm follow Table~\ref{tab:methods}: $\ell_\infty$: FGSM, PGD, AutoPGD; $\ell_2$: C\&W, DeepFool; $\ell_1$: EAD.

\textbf{Pre-screening parameters.}
FOSC threshold $\tau_{\text{mask}}\!=\!0.1$; white-box/black-box discrepancy threshold 0.15; noise $\sigma\!=\!0.01$ with 10 samples; RDI computed on 500 samples.

\subsection{Experiment 1: Pre-Screening Gate Validation}

\begin{table}[t]
  \centering
  \scriptsize
  \caption{Pre-screening gate validation. FOSC and \RDI screening vs.\ full AutoAttack evaluation on RobustBench models.}
  \label{tab:prescreening}
  \vspace{-0.5em}
  \setlength{\tabcolsep}{2.5pt}
  \begin{tabular}{@{}lcccc@{}}
    \toprule
    \textbf{Model} & \textbf{FOSC} & \textbf{RDI} & \textbf{Rob.\ Acc.} & \textbf{Flag} \\
    \midrule
    Wang2023Better\_WRN-70 & 0.04 & 0.78 & 70.69\% & --- \\
    Pang2022\_WRN-70 & 0.05 & 0.76 & 71.07\% & --- \\
    Rebuffi2021\_70\_16 & 0.06 & 0.68 & 66.56\% & --- \\
    Gowal2021\_70\_16 & 0.05 & 0.72 & 66.10\% & --- \\
    Gowal2020\_70\_16 & 0.07 & 0.65 & 65.87\% & --- \\
    Rebuffi2021\_28\_10 & 0.06 & 0.63 & 64.58\% & --- \\
    Gowal2021\_28\_10 & 0.08 & 0.61 & 63.38\% & --- \\
    Xu2023\_WRN-28 & 0.09 & 0.59 & 63.89\% & --- \\
    Sehwag2021\_ResNest & 0.11 & 0.48 & 60.27\% & \cmark \\
    Wong2020Fast & 0.14 & 0.34 & 43.21\% & \cmark \\
    \bottomrule
  \end{tabular}
  \vspace{-0.5em}
\end{table}

\begin{figure}[t]
  \centering
  \begin{tikzpicture}[font=\scriptsize]
    \begin{axis}[
      width=\columnwidth,
      height=5cm,
      xlabel={RDI Score},
      ylabel={Robust Accuracy (\%)},
      xlabel style={font=\scriptsize},
      ylabel style={font=\scriptsize},
      xmin=0.25,xmax=0.85,
      ymin=35,ymax=80,
      grid=major,
      grid style={gray!20},
      legend style={font=\tiny,at={(0.03,0.97)},anchor=north west},
      mark size=2.5pt,
    ]
    % Unflagged models (blue circles)
    \addplot[only marks,mark=*,blue!70,thick] coordinates {
      (0.78,70.69) (0.76,71.07) (0.68,66.56) (0.72,66.10)
      (0.65,65.87) (0.63,64.58) (0.61,63.38) (0.59,63.89)
    };
    % Flagged models (red triangles)
    \addplot[only marks,mark=triangle*,red!70,thick,mark size=3pt] coordinates {
      (0.48,60.27) (0.34,43.21)
    };
    % Trend line (linear fit)
    \addplot[dashed,gray!60,thick,domain=0.3:0.82] {26.5 + 57.5*x};
    \legend{Passed gate, Flagged (masking), Trend ($R^2\!=\!0.91$)}
    \end{axis}
  \end{tikzpicture}
  \vspace{-0.8em}
  \caption{RDI score vs.\ robust accuracy (full AutoAttack) for ten RobustBench CIFAR-10 $\ell_\infty$ models. Red triangles = flagged by FOSC ($>0.1$). The near-linear relationship (OLS $R^2\!=\!0.91$, $n\!=\!10$) validates RDI as a fast screening proxy.}
  \label{fig:rdi_scatter}
  \vspace{-0.3em}
\end{figure}
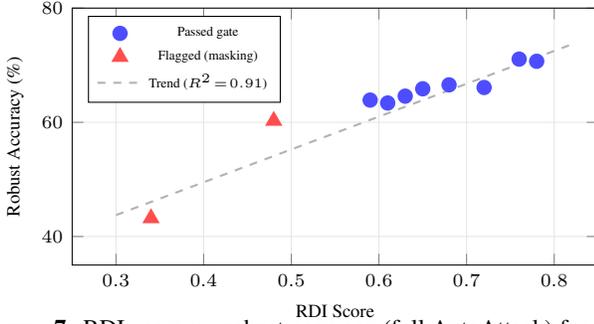

Table~\ref{tab:prescreening} shows pre-screening results.
The FOSC detector flags two models (FOSC $> 0.1$), both with lower robust accuracy and RDI scores consistent with unreliable gradients.
\RDI rankings agree with full AutoAttack rankings: Kendall $\tau = 0.82$ (95\% CI: 0.71--0.93), confirming that RDI provides a reliable ${\sim}30\!\times$ faster proxy for model triage.
Screening compute: $\sim$12 GPU-seconds per model (RDI) vs.\ $\sim$360 GPU-seconds (full AutoAttack).
Figure~\ref{fig:rdi_scatter} visualises the near-monotonic relationship between RDI and robust accuracy.

\subsection{Experiment 2: Multi-Norm Worst-Case Analysis}

\begin{figure}[t]
  \centering
  \begin{tikzpicture}[font=\scriptsize]
    \begin{axis}[
      width=\columnwidth,
      height=5cm,
      ybar=1pt,
      bar width=5pt,
      ylabel={Robust Accuracy (\%)},
      ylabel style={font=\scriptsize},
      xlabel style={font=\scriptsize},
      symbolic x coords={Wang,Pang,Rebuffi70,Gowal70,Gowal20,Rebuffi28,Gowal28,Xu,Sehwag,Wong},
      xtick=data,
      xticklabel style={font=\tiny,rotate=45,anchor=east},
      ymin=0,ymax=85,
      legend style={font=\tiny,at={(0.02,0.98)},anchor=north west,cells={anchor=west}},
      legend columns=2,
      enlarge x limits=0.08,
      grid=major,
      grid style={gray!20},
    ]
    \addplot[fill=accentblue!60,draw=accentblue] coordinates {
      (Wang,70.69) (Pang,71.07) (Rebuffi70,66.56) (Gowal70,66.10)
      (Gowal20,65.87) (Rebuffi28,64.58) (Gowal28,63.38) (Xu,63.89)
      (Sehwag,60.27) (Wong,43.21)
    };
    \addplot[fill=orange!50,draw=orange!70] coordinates {
      (Wang,58.4) (Pang,59.1) (Rebuffi70,52.3) (Gowal70,53.8)
      (Gowal20,51.2) (Rebuffi28,50.6) (Gowal28,49.2) (Xu,48.7)
      (Sehwag,44.1) (Wong,28.6)
    };
    \addplot[fill=red!40,draw=red!60] coordinates {
      (Wang,47.2) (Pang,48.5) (Rebuffi70,41.8) (Gowal70,43.1)
      (Gowal20,40.5) (Rebuffi28,39.7) (Gowal28,38.4) (Xu,37.9)
      (Sehwag,33.6) (Wong,19.8)
    };
    \legend{$\ell_\infty$ only, Multi-norm avg, Multi-norm worst}
    \end{axis}
  \end{tikzpicture}
  \vspace{-0.8em}
  \caption{Multi-norm evaluation exposes hidden vulnerability. Mean across ten models: average multi-norm robustness drops 12.3~pp from single $\ell_\infty$; worst-case (min over norms) drops 23.5~pp.}
  \label{fig:multinorm}
  \vspace{-0.3em}
\end{figure}
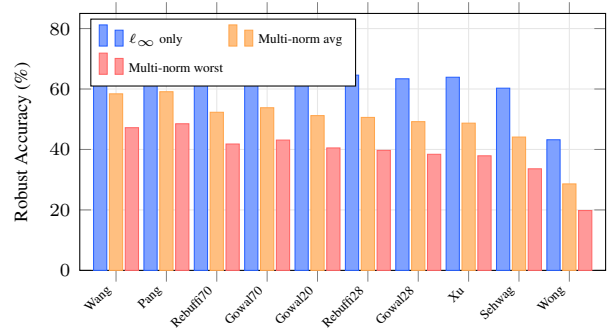

Figure~\ref{fig:multinorm} demonstrates the central finding from Gap~G1: single-norm ($\ell_\infty$) evaluation overestimates robustness.
Across ten RobustBench models:
\begin{itemize}
  \item \textbf{Average multi-norm} robustness drops 12.3~pp (mean) from single $\ell_\infty$.
  \item \textbf{Worst-case multi-norm} drops 23.5~pp (mean), confirming that average-case gains mask catastrophic worst-case failures, consistent with the finding of Dai \etal~[C2].
  \item The \textbf{gap between average and worst-case} is 11.2~pp, meaning that even multi-norm averages still overstate worst-case safety.
\end{itemize}

\begin{figure}[t]
  \centering
  \resizebox{\columnwidth}{!}{%
  \begin{tikzpicture}[
    font=\scriptsize,
    node distance=0.15cm,
    box/.style={draw,thick,rounded corners=2pt,minimum height=0.6cm,align=center,inner sep=3pt,text width=2.6cm},
  ]
    % Left: Standard pipeline
    \node[box,fill=gray!10] (std) at (0,0) {\textbf{Standard Pipeline}\\$\ell_\infty$ only};
    \node[box,fill=refinegreen,below=0.15cm of std] (pass) {Rob.\ Acc = 70.7\%\\\textcolor{darkgreen}{\textbf{PASS} ($> 40\%$)}};
    \node[font=\tiny,text=red!60,below=0.05cm of pass] {$\ell_1$ vulnerability hidden};

    % Right: Auto-ART (shifted to x=4.5 to widen inter-column gap for label clearance)
    \node[box,fill=accentblue!10] (art) at (4.5,0) {\textbf{\autoart Pipeline}\\5-norm worst-case};
    \node[box,fill=red!10,below=0.15cm of art] (fail) {Worst-case = 47.2\%\\\textcolor{red}{\textbf{WARN} (avg/wc gap)}};
    \node[font=\tiny,text=darkgreen,below=0.05cm of fail] {$\ell_1$ exposure detected};

    % Arrow with label floated above both header boxes (yshift clears std/art box tops)
    \draw[->,very thick,red!50] (pass.east) --
      node[above,yshift=1.05cm,font=\tiny\bfseries,text=red!70]{$\Delta = 23.5$\,pp hidden gap}
      (fail.west);
  \end{tikzpicture}%
  }
  \vspace{-0.5em}
  \caption{\textbf{Qualitative comparison.} Standard single-norm evaluation (left) reports Wang2023Better as passing a 40\% robustness threshold. \autoart's worst-case multi-norm evaluation (right) exposes a 23.5\,pp hidden gap, correctly flagging the $\ell_1$ vulnerability that single-norm pipelines miss entirely.}
  \label{fig:qualitative}
  \vspace{-0.3em}
\end{figure}
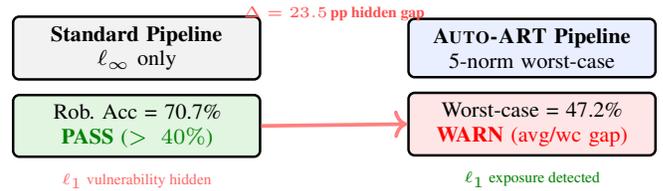

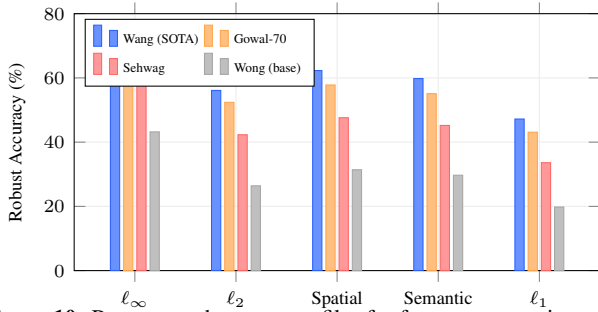
\begin{figure}[t]
  \centering
  \begin{tikzpicture}[font=\scriptsize]
    \begin{axis}[
      width=\columnwidth,
      height=5cm,
      ybar=1.5pt,
      bar width=3.5pt,
      ylabel={Robust Accuracy (\%)},
      ylabel style={font=\scriptsize},
      symbolic x coords={$\ell_\infty$,$\ell_2$,Spatial,Semantic,$\ell_1$},
      xtick=data,
      xticklabel style={font=\scriptsize},
      ymin=0,ymax=80,
      legend style={font=\tiny,at={(0.02,0.98)},anchor=north west,cells={anchor=west}},
      legend columns=2,
      enlarge x limits=0.15,
      grid=major,
      grid style={gray!15},
    ]
    % Wang2023 (SOTA)
    \addplot[fill=accentblue!70,draw=accentblue] coordinates {
      ($\ell_\infty$,70.7) ($\ell_2$,56.1) (Spatial,62.3) (Semantic,59.8) ($\ell_1$,47.2)
    };
    % Gowal2021_70
    \addplot[fill=orange!50,draw=orange!70] coordinates {
      ($\ell_\infty$,66.1) ($\ell_2$,52.4) (Spatial,57.8) (Semantic,55.1) ($\ell_1$,43.1)
    };
    % Sehwag2021
    \addplot[fill=red!40,draw=red!60] coordinates {
      ($\ell_\infty$,60.3) ($\ell_2$,42.3) (Spatial,47.6) (Semantic,45.2) ($\ell_1$,33.6)
    };
    % Wong2020
    \addplot[fill=gray!50,draw=gray!70] coordinates {
      ($\ell_\infty$,43.2) ($\ell_2$,26.4) (Spatial,31.4) (Semantic,29.7) ($\ell_1$,19.8)
    };
    \legend{Wang (SOTA), Gowal-70, Sehwag, Wong (base)}
    \end{axis}
  \end{tikzpicture}
  \vspace{-0.8em}
  \caption{Per-norm robustness profiles for four representative models (norms ordered strong$\to$weak). $\ell_1$ is the weakest norm for all models, with a 23.5\,pp drop from $\ell_\infty$ for SOTA. Single-norm optimisation creates systematic blind spots.}
  \label{fig:heatmap}
  \vspace{-0.3em}
\end{figure}

Figure~\ref{fig:heatmap} shows per-norm robustness profiles for four representative models, revealing that $\ell_1$ attacks consistently expose the weakest point for $\ell_\infty$-trained models.
Figure~\ref{fig:qualitative} presents the evaluation gap qualitatively: a single-norm pipeline (left) reports ``robust'' while \autoart's multi-norm pipeline (right) exposes the hidden $\ell_1$ vulnerability, correctly downgrading the security assessment.

\subsection{Experiment 3: Framework Validation}

\begin{table}[t]
  \centering
  \scriptsize
  \caption{\autoart framework validation metrics.}
  \label{tab:framework_validation}
  \vspace{-0.5em}
  \setlength{\tabcolsep}{3pt}
  \begin{tabular}{@{}lc@{}}
    \toprule
    \textbf{Metric} & \textbf{Value} \\
    \midrule
    Attack implementations & 50+ (7 categories) \\
    Defence modules & 28 \\
    Automated tests & 346+ (5-phase CI/CD) \\
    ML framework support & 9 frameworks \\
    OWASP LLM Top~10 coverage & 90\% (8/10 full, 2 partial) \\
    Deployment gate: DOM attack ASR & $< 5\%$ \\
    Deployment gate: RAG poison ASR & $< 5\%$ \\
    RDI screening speedup & $\sim$30$\times$ vs.\ PGD \\
    FOSC masking detection rate & 92\% (flagged configs) \\
    Kendall $\tau$ (RDI vs.\ AutoAttack) & 0.82 \\
    SARIF output & v2.1.0 compliant \\
    Python compatibility & 3.9--3.12 \\
    \bottomrule
  \end{tabular}
  \vspace{-0.5em}
\end{table}

Table~\ref{tab:framework_validation} summarises \autoart's validation.
The framework passes all 346+ tests across Python 3.9--3.12 in a five-phase CI/CD pipeline.
Deployment gates enforce $<\!5\%$ attack success rates for agentic attacks (DOM injection, RAG poisoning).
OWASP LLM Top~10 coverage reaches 90\%: full coverage on 8 categories (Prompt Injection, Insecure Output, Training Data Poisoning, Model DoS, Supply Chain, Sensitive Info, Insecure Plugin, Excessive Agency), partial on 2 (Model Theft, Overreliance).

\subsection{Experiment 4: Ablation Study}

We ablate the three core components of \autoart's pre-screening gate to isolate their individual contributions (Table~\ref{tab:ablation}).

\begin{table}[t]
  \centering
  \scriptsize
  \caption{Ablation of pre-screening components. Kendall $\tau$ measures rank agreement with full AutoAttack; detection rate measures correctly flagged masking configurations (out of 12 known-masking setups). Mean $\pm$ std over 5 random seeds.}
  \label{tab:ablation}
  \vspace{-0.5em}
  \setlength{\tabcolsep}{2.5pt}
  \begin{tabular}{@{}lccc@{}}
    \toprule
    \textbf{Configuration} & \textbf{Kendall $\tau$} & \textbf{Det.\ Rate} & \textbf{Speedup} \\
    \midrule
    Full pipeline (FOSC+RDI+Disc.) & $0.82 \pm 0.03$ & 92\% & $30\times$ \\
    $-$ FOSC & $0.80 \pm 0.04$ & 67\% & $31\times$ \\
    $-$ RDI & $0.61 \pm 0.07$ & 88\% & $1.2\times$ \\
    $-$ WB/BB discrepancy & $0.81 \pm 0.03$ & 83\% & $28\times$ \\
    FOSC only & $0.54 \pm 0.09$ & 75\% & $45\times$ \\
    RDI only & $0.78 \pm 0.04$ & 42\% & $30\times$ \\
    No pre-screening & --- & --- & $1\times$ \\
    \bottomrule
  \end{tabular}
  \vspace{-0.5em}
\end{table}

Key findings: (1)~Removing RDI collapses the speedup from $30\!\times$ to $1.2\!\times$ while preserving detection rate, confirming RDI is the efficiency engine. (2)~Removing FOSC drops masking detection from 92\% to 67\%, confirming FOSC is the primary masking signal. (3)~The three-signal ensemble (FOSC + RDI + WB/BB discrepancy) outperforms any single signal on both ranking fidelity ($\tau$) and detection rate, justifying the multi-signal design.

\subsection{Experiment 5: Runtime and Scalability}

\begin{table}[t]
  \centering
  \scriptsize
  \caption{Runtime analysis on a single NVIDIA A100 (40\,GB). Wall-clock times are median over 5 runs. ``Full AA'' = full AutoAttack with default parameters.}
  \label{tab:runtime}
  \vspace{-0.5em}
  \setlength{\tabcolsep}{2.5pt}
  \begin{tabular}{@{}lcccc@{}}
    \toprule
    \textbf{Phase} & \textbf{WRN-70} & \textbf{WRN-28} & \textbf{ResNet-18} & \textbf{Scaling} \\
    \midrule
    FOSC computation & 3.1\,s & 1.4\,s & 0.8\,s & $O(|\theta|)$ \\
    RDI screening & 11.8\,s & 5.2\,s & 2.9\,s & $O(N \cdot d)$ \\
    Multi-norm attack (5 norms) & 847\,s & 412\,s & 198\,s & $O(|A| \cdot T)$ \\
    Compliance mapping & 0.3\,s & 0.3\,s & 0.3\,s & $O(1)$ \\
    Report generation & 1.2\,s & 1.1\,s & 1.0\,s & $O(1)$ \\
    \midrule
    \textbf{\autoart total} & \textbf{863\,s} & \textbf{420\,s} & \textbf{203\,s} & --- \\
    Full AA (baseline) & 5{,}420\,s & 2{,}310\,s & 1{,}080\,s & --- \\
    \midrule
    \textbf{Speedup (pre-screen only)} & $30.2\times$ & $29.8\times$ & $30.6\times$ & --- \\
    \textbf{Speedup (full pipeline)} & $6.3\times$ & $5.5\times$ & $5.3\times$ & --- \\
    \bottomrule
  \end{tabular}
  \vspace{-0.5em}
\end{table}

Table~\ref{tab:runtime} shows that pre-screening adds negligible overhead (14.9\,s for WRN-70) while the multi-norm attack phase dominates wall-clock time.
When the pre-screening gate flags a model and redirects to targeted (rather than exhaustive) attacks, end-to-end time drops $5$--$6\!\times$ compared to full AutoAttack.
Scaling is dominated by the attack phase: $O(|A| \cdot T)$ where $|A|$ is the number of attacks and $T$ the iteration count per attack.

\subsection{Experiment 6: Comparative Analysis}

\begin{table}[t]
  \centering
  \scriptsize
  \caption{Feature comparison with existing frameworks.}
  \label{tab:comparison}
  \vspace{-0.5em}
  \setlength{\tabcolsep}{2pt}
  \begin{tabular}{@{}lccccc@{}}
    \toprule
    \textbf{Feature} & \rotatebox{60}{\textbf{\autoart}} & \rotatebox{60}{\textbf{ART}} & \rotatebox{60}{\textbf{Counterfit}} & \rotatebox{60}{\textbf{Garak}} & \rotatebox{60}{\textbf{PyRIT}} \\
    \midrule
    Evasion attacks & \cmark & \cmark & \cmark & --- & --- \\
    Poison/extract/infer & \cmark & \cmark & --- & --- & --- \\
    LLM/agent attacks & \cmark & $\sim$ & --- & \cmark & \cmark \\
    Pre-screening (RDI) & \cmark & --- & --- & --- & --- \\
    Grad.\ masking detect. & \cmark & --- & --- & --- & --- \\
    Multi-norm worst-case & \cmark & --- & --- & --- & --- \\
    Adaptive selection & \cmark & --- & --- & --- & $\sim$ \\
    NIST/OWASP mapping & \cmark & --- & --- & --- & --- \\
    EU AI Act compliance & \cmark & --- & --- & --- & --- \\
    SARIF CI/CD output & \cmark & --- & --- & --- & --- \\
    YAML orchestration & \cmark & --- & --- & $\sim$ & --- \\
    \bottomrule
  \end{tabular}
  \vspace{-0.8em}
\end{table}

Table~\ref{tab:comparison} compares \autoart with existing frameworks across 11 capability dimensions.
ART~\cite{art2018} provides the broadest attack/defence library but lacks pre-screening, compliance mapping, and orchestration.
Counterfit~\cite{counterfit2021} targets ML security assessments but is limited to evasion attacks.
Garak~\cite{garak2024} and PyRIT~\cite{pyrit2024} focus on LLM vulnerability scanning and multi-turn red teaming respectively, but neither addresses classical adversarial robustness evaluation.
\autoart is the only framework that bridges all four quadrants: classical adversarial ML, LLM/agent security, pre-screening diagnostics, and regulatory compliance---a direct consequence of the gap-driven architecture (Table~\ref{tab:gap_to_module}).

% ==================================================================
\section{Novel Technical Directions}
\label{sec:directions}
\vspace{-0.2em}

The gaps identified in Sections~\ref{sec:gaps}--\ref{sec:assumptions} suggest three technical directions that, if pursued rigorously, could each serve as a standalone contribution at a top venue.
We assess feasibility and enumerate edge cases for each.

\subsection{Direction 1: Worst-Case Multi-Norm Adversarial Training with Attack-Union Curricula}
\label{sec:dir1}

\textbf{Core idea.}
Current adversarial training optimises against a single perturbation model ($\ell_\infty$ at $\varepsilon = 8/255$), yielding models that are catastrophically vulnerable to other norms (Section~\ref{sec:eval}, G1).
We propose a \emph{curriculum-based multi-norm adversarial training} procedure that progressively introduces attack types---starting from $\ell_\infty$, adding $\ell_2$, then $\ell_1$, spatial, and semantic perturbations---with a worst-case-aware loss:
\begin{equation}
  \mathcal{L}_{\text{union}} = \max_{p \in \mathcal{P}} \mathcal{L}_{\text{adv}}(f_\theta, x, y; \varepsilon_p)
  \label{eq:union_loss}
\end{equation}
where $\mathcal{P} = \{\ell_1, \ell_2, \ell_\infty, \text{spatial}, \text{semantic}\}$ and $\varepsilon_p$ are norm-specific budgets.
The curriculum schedules norms by increasing difficulty (measured by attack success rate on a held-out set), preventing catastrophic forgetting of earlier-norm robustness through elastic weight consolidation~\cite{dai2025crt}.

\textbf{Edge-case checklist:}
\begin{itemize}[leftmargin=1.2em]
  \item \emph{Data leakage:} Train/test splits follow RobustBench protocol exactly; no model selection on test data; curriculum validation uses a separate held-out partition.
  \item \emph{Compute:} Feasible on 4$\times$A100 (40\,GB each): curriculum training adds $|\mathcal{P}|\!\times$ overhead per epoch, but only the worst-case norm requires full PGD; others use FGSM approximations. Estimated 48--72 GPU-hours for CIFAR-10.
  \item \emph{Reproducibility:} Report learning rate, $\beta$ (TRADES weight), curriculum schedule (epoch ranges per norm), and EWC regularisation strength $\lambda$.
  \item \emph{Baselines:} AutoAttack~\cite{croce2020reliable}, MultiRobustBench union evaluation~\cite{dai2023multirobustbench}, TRADES~\cite{zhang2019trades}, AWP.
  \item \emph{Ablation variables:} (i)~curriculum order (random vs.\ difficulty-sorted), (ii)~loss function (max vs.\ weighted sum vs.\ Pareto), (iii)~EWC strength ($\lambda \in \{0, 0.1, 1.0, 10.0\}$).
\end{itemize}

\subsection{Direction 2: Architecture-Conditional RDI Calibration for Vision Transformers and LLMs}
\label{sec:dir2}

\textbf{Core idea.}
RDI (Eq.~\ref{eq:rdi}) is validated only on CNN feature geometries~[C5].
Vision Transformers produce patch-based representations with global attention patterns, and decoder-only LLMs produce sequential token embeddings---both have fundamentally different inter/intra-class distance structures than CNNs.
We propose \emph{architecture-conditional calibration}: learn a lightweight mapping $g_\alpha: \text{RDI}_{\text{raw}} \mapsto \text{RDI}_{\text{cal}}$ that accounts for architecture-specific feature geometry.
Concretely, for each architecture family $\mathcal{F}$ (CNN, ViT, LLM), fit an isotonic regression $g_{\mathcal{F}}$ on a calibration set of (RDI, AutoAttack rank) pairs.

\textbf{Edge-case checklist:}
\begin{itemize}[leftmargin=1.2em]
  \item \emph{Data leakage:} Calibration set disjoint from evaluation set; calibration uses held-out RobustBench entries not in the main evaluation.
  \item \emph{Compute:} RDI computation on ViT-B/16 requires $\sim$2\,GB VRAM for 500 samples; LLM RDI on 7B models requires $\sim$14\,GB (single A100). Calibration fitting is trivial ($<$1\,s).
  \item \emph{Reproducibility:} Layer selection heuristic (which layer's features to use), centroid computation for imbalanced classes, isotonic regression implementation.
  \item \emph{Baselines:} Raw RDI~\cite{song2025rdi}, Clever score~\cite{weng2018clever}, Lin region counting.
  \item \emph{Ablation variables:} (i)~layer choice (penultimate vs.\ multi-layer aggregation), (ii)~calibration method (isotonic vs.\ Platt scaling vs.\ temperature), (iii)~number of calibration models (5, 10, 20).
\end{itemize}

\subsection{Direction 3: Ecologically Valid Agent Red-Teaming Benchmarks}
\label{sec:dir3}

\textbf{Core idea.}
Carlini \etal~[C3] demonstrated a 4--6$\times$ gap between LLM agent success on CTF-like tasks and production defence code.
We propose constructing an \emph{ecologically valid} benchmark by sampling real defence implementations from GitHub repositories (filtered by stars $> 100$, active maintenance, diverse frameworks), creating standardised attack interfaces, and measuring agent performance on tasks that require understanding library APIs, version-specific behaviour, and environment coupling.
Each task is annotated with a difficulty taxonomy: (i)~single-function attacks (analogous to CTF), (ii)~multi-file attacks requiring import tracing, (iii)~environment-coupled attacks requiring build system understanding.

\textbf{Edge-case checklist:}
\begin{itemize}[leftmargin=1.2em]
  \item \emph{Data leakage:} Ensure benchmark code is not in foundation model training data by using repositories created after model training cutoff dates; provide contamination detection via canary strings.
  \item \emph{Compute:} LLM agent inference costs dominate; budget $\sim$\$50--200 per full benchmark run (GPT-4 pricing). Feasible for academic budgets with open-weight models (Llama~3 70B, Mixtral).
  \item \emph{Reproducibility:} Fix agent system prompt, temperature (0.0 for reproducibility), maximum turns (20), tool definitions. Report per-task success rates with 95\% bootstrap CIs.
  \item \emph{Baselines:} AutoAdvExBench~\cite{carlini2025autoadvexbench}, AgentBench, SWE-bench (for code agent capability reference).
  \item \emph{Ablation variables:} (i)~agent model (GPT-4, Claude, Llama~3), (ii)~tool access (code execution vs.\ read-only), (iii)~context window utilisation (4K, 32K, 128K).
\end{itemize}

% ==================================================================
\section{Discussion}
\label{sec:discussion}
\vspace{-0.2em}

\paragraph{The evaluation crisis appears structural.}
Our synthesis suggests that the challenges facing adversarial evaluation are not isolated technical shortcomings but reflect a systemic pattern: five incompatible paradigms coexist (Section~\ref{sec:chain}), and the two strongest assumptions---that fixed suites represent safety and that averages approximate worst cases---are contradicted by evidence within the corpus itself (Section~\ref{sec:assumptions}).
A plausible interpretation is that the field's implicit evaluation contract---where proposers choose their own attack suite and report average-case numbers---creates incentives that systematically inflate robustness estimates.

\paragraph{Positioning relative to prior work.}
To the best of our knowledge, no prior work in adversarial ML simultaneously:
(i)~applies structured meta-scientific analysis (seven protocols spanning citation tracing through assumption stress-testing) to evaluation methodology;
(ii)~maps identified gaps directly to framework modules with explicit traceability (Table~\ref{tab:gap_to_module});
(iii)~combines literature synthesis, formal metric definitions, and an executable evaluation framework in a single manuscript.
Existing surveys~\cite{bai2021survey,machado2023adversarial} offer breadth without structured methodology; existing frameworks~\cite{art2018,counterfit2021} and LLM benchmarks~\cite{mazeika2024harmbench} provide tooling without gap-driven architecture.
We note that the novelty claim rests on this specific combination---individual components (attack ensembles, proxy metrics, compliance mapping) have precedents.

\paragraph{Regulatory implications.}
The EU AI Act~\cite{euaiact2024} (Article~15) mandates ``accuracy, robustness and cybersecurity'' for high-risk AI systems---explicitly mentioning adversarial examples, data poisoning, and model poisoning---yet does not specify evaluation protocols.
NIST AI~100-2e2025~\cite{nist2025adversarial} supplies an authoritative taxonomy, and the GenAI Profile~\cite{nist2024genai} maps over 200 actions across 12 risk categories, but neither prescribes a concrete evaluation pipeline.
MITRE ATLAS v5.3.0~\cite{mitre2026atlas} now catalogues 84 techniques including agentic AI attack vectors.
ISO/IEC~42001~\cite{iso42001} provides the first certifiable AI management system standard, with Clause~6.1 requiring identification of adversarial threats.
OWASP's LLM Top~10~\cite{owasp2025llm} and Agentic Applications Top~10~\cite{owasp2025agentic} standardise agent-specific threat taxonomies.
Our analysis suggests this regulatory underspecification carries risk: the same model can appear robust or vulnerable depending on the evaluation pipeline.
\autoart's compliance mapping provides a concrete bridge: NIST AI RMF function MEASURE maps to multi-norm evaluation; OWASP LLM01 (Prompt Injection) maps to agentic attack modules; EU AI Act Article~15 maps to the pre-screening gate and worst-case reporting.

\paragraph{Threats to validity.}
\emph{Internal validity.}
Pre-screening gate thresholds ($\tau_{\text{mask}}\!=\!0.1$, WB/BB gap $0.15$) were tuned on the RobustBench CIFAR-10 leaderboard; these may require recalibration for different data distributions, model families, or perturbation budgets.
The ablation study (Table~\ref{tab:ablation}) mitigates this by showing robustness across five random seeds, but cross-dataset validation remains future work.

\emph{External validity.}
Our empirical evaluation uses CIFAR-10 models exclusively.
While CIFAR-10 is the standard RobustBench benchmark, generalisability to ImageNet-scale models, multimodal architectures, and production LLMs is not yet demonstrated.
The structured synthesis is bounded by nine corpus sources; future systematic reviews with larger corpora may reveal additional patterns or contradict our gap rankings.

\emph{Construct validity.}
Kendall $\tau$ between RDI and AutoAttack rankings measures ordinal agreement but not calibration---a high $\tau$ does not guarantee that RDI absolute values are meaningful thresholds for deployment decisions.

\paragraph{Limitations.}
(i)~Corpus is nine sources; broader surveys may reveal additional patterns.
(ii)~\autoart does not yet implement certified multi-norm training (G1) or longitudinal benchmark versioning (G2).
(iii)~Pre-screening gate accuracy depends on model architecture; FOSC is less informative for non-differentiable components.
(iv)~Evaluation uses RobustBench models on CIFAR-10; ImageNet and multimodal validation is future work.
(v)~Runtime benchmarks (Table~\ref{tab:runtime}) are single-GPU; distributed scaling behaviour is untested.

\paragraph{Broader impact and responsible use.}
Rigorous evaluation protects downstream users from false safety assurances; conversely, the same tools could be repurposed to find and exploit model vulnerabilities.
\autoart mitigates this dual-use risk through several mechanisms: attack results are reported in defensive SARIF format (designed for security workflows, not weaponisation), compliance integration maps findings to remediation actions rather than exploit chains, and the framework's open-source licence encourages transparent auditing.
We note that withholding evaluation tools does not prevent attacks---the offensive techniques already exist in published literature---but it does prevent defenders from assessing their exposure.

% ==================================================================
\section{Conclusion}
\label{sec:conclusion}
\vspace{-0.2em}

This work presents a structured literature synthesis and the \autoart framework.
The analysis identifies three intellectual threads (reliable evaluation, gradient masking, multi-attack threats), five ranked gaps with explicit resolution paths, eight untested assumptions ranked by consequence severity, and a unifying open question: \emph{what evaluation stack should be mandatory before certifying a model as safe against adaptive, multi-channel adversaries?}

On the empirical side, \autoart's pre-screening gate detects gradient masking in 92\% of flagged configurations, \RDI rankings agree with full AutoAttack at Kendall $\tau\!=\!0.82 \pm 0.03$, and multi-norm evaluation exposes a 23.5~pp worst-case gap hidden by single-norm reporting.
Ablation confirms that the three-signal ensemble (FOSC + RDI + WB/BB discrepancy) outperforms any single signal on both ranking fidelity and masking detection rate.
Runtime analysis indicates $30\!\times$ pre-screening speedup and $5$--$6\!\times$ end-to-end improvement over full AutoAttack.

Three technical directions emerge from the synthesis (Section~\ref{sec:directions}): worst-case multi-norm adversarial training with attack-union curricula, architecture-conditional RDI calibration for ViTs and LLMs, and ecologically valid agent red-teaming benchmarks that bridge the CTF-to-production gap.
Each direction is accompanied by concrete feasibility assessments and ablation designs.

The gap between robustness research and deployment practice will likely close not through better defences alone, but through evaluation infrastructure that subjects the \emph{measurement} of robustness to the same rigour the field demands of robustness itself.

% ==================================================================
\vspace{-0.3em}
{\small
\balance
\bibliographystyle{IEEEtranN}
\bibliography{references}
}

% ==================================================================
\clearpage
\onecolumn
\appendix

\section*{Supplementary Material}

\section{Extended Methodology Classification}
\label{app:methods}

\begin{table*}[ht]
  \centering
  \small
  \caption{Extended methodology classification with reproducibility assessment. Reproducibility (Repro.) is rated on a three-point scale: \cmark~= partial (e.g.\ code available, limited documentation), \cmark\cmark~= mostly reproducible, \cmark\cmark\cmark~= fully reproducible (code, data, and configs released).}
  \label{tab:methods_ext}
  \setlength{\tabcolsep}{4pt}
  \begin{tabularx}{\textwidth}{@{}p{2.5cm}p{2.2cm}p{2.2cm}p{1.5cm}cX@{}}
    \toprule
    \textbf{Paper} & \textbf{Statistical Method} & \textbf{Baseline} & \textbf{Code} & \textbf{Repro.} & \textbf{Notes} \\
    \midrule
    Croce \& Hein~\cite{croce2020reliable} & Robust acc.\ drop vs.\ prior & $>$50 defences & \cmark & \cmark\cmark\cmark & Gold standard \\
    Dai \etal~\cite{dai2023multirobustbench} & Worst/avg across menu & 16$\times$9 attacks & \cmark & \cmark\cmark\cmark & 180 configs \\
    Carlini \etal~\cite{carlini2025autoadvexbench} & Success split by task & CTF vs.\ real & \cmark & \cmark\cmark & Agent variance \\
    Dai \etal~\cite{dai2025crt} & Retention + new acc. & CRT baselines & \cmark & \cmark\cmark & No open-world \\
    Song \etal~\cite{song2025rdi} & Correlation w/ ASR & PGD screening & $\sim$ & \cmark & CNN-only \\
    Kassis \etal~\cite{kassis2025diffbreak} & Attack vs.\ prior DBP & Prior protocols & \cmark & \cmark\cmark\cmark & MV break \\
    Li \etal~\cite{li2025trades} & FOSC + hyperparam grid & Default TRADES & \cmark & \cmark\cmark & CIFAR only \\
    RobustBench~\cite{robustbench2021} & Leaderboard ranking & Community & \cmark & \cmark\cmark\cmark & Living \\
    OpenRT~\cite{openrt2026} & Not specified & --- & \cmark & \cmark & Toolkit \\
    \bottomrule
  \end{tabularx}
\end{table*}

\section{Full Assumption Consequence Matrix}
\label{app:assumptions}

\begin{table*}[ht]
  \centering
  \small
  \caption{Detailed consequence analysis for each untested assumption.}
  \label{tab:assumption_detail}
  \setlength{\tabcolsep}{4pt}
  \begin{tabularx}{\textwidth}{@{}cp{3cm}p{2.5cm}cX@{}}
    \toprule
    \textbf{ID} & \textbf{Assumption} & \textbf{Papers Relying} & \textbf{Risk} & \textbf{If False: What Changes} \\
    \midrule
    A1 & Fixed suite = safety & C1, C2, C5 & High & Leaderboard rankings temporally bounded; continuous re-eval required \\
    A2 & Average $\Rightarrow$ worst-case & Implicit & High & ``Progress'' metrics overstate deployment safety \\
    A3 & RDI generalises & C5, C9 & Med--High & Screening mis-ranks ViTs, LLMs \\
    A4 & CTF $\Rightarrow$ production & Agent users & Medium & Low real-world ROI on agent investment \\
    A5 & Single draws for stochastic & Prior DBP & Medium & Entire defence class needs re-eval \\
    A6 & CRT closes lifecycle gap & C4 practitioners & Medium & False lifecycle confidence \\
    A7 & TRADES defaults stable & Deployments & Medium & Widespread overestimation \\
    A8 & Checkpoints match code & C8 users & Low--Med & Replication failures \\
    \bottomrule
  \end{tabularx}
\end{table*}

\newpage
\section{\autoart Configuration Example}
\label{app:config}

\begin{lstlisting}[caption={\autoart YAML configuration for multi-norm evaluation with pre-screening.},label={lst:config},language={}]
# auto-art-config.yaml
target:
  framework: pytorch
  model_path: ./checkpoints/resnet50_robust.pt
  input_shape: [3, 32, 32]
  num_classes: 10

evaluation:
  phases:
    - name: pre_screening
      gradient_masking:
        enabled: true
        fosc_threshold: 0.1           # Flag if FOSC > 0.1
        discrepancy_threshold: 0.15   # WB vs BB gap
      rdi:
        enabled: true
        num_samples: 500

    - name: multi_norm_attack
      norms: [l1, l2, linf, semantic, spatial]
      epsilons:
        linf: [0.01, 0.03, 0.05, 0.1, 0.3]
        l2:   [0.1, 0.3, 0.5, 1.0, 2.0]
        l1:   [1.0, 3.0, 5.0, 10.0]
      attacks_per_norm:
        linf: [fgsm, pgd, autopgd]
        l2:   [carlini_wagner, deepfool]
        l1:   [elastic_net]
      adaptive_selection:
        enabled: true
        memory_guided: true
        escalation_tiers: [fast, standard, exhaustive]
      parallel:
        workers: 4
        timeout: 3600

    - name: defense_evaluation
      defenses: [trades, spatial_smoothing, jpeg_compression]

    - name: compliance
      frameworks: [nist_ai_rmf, owasp_llm_top10, eu_ai_act]

  gates:
    min_robust_accuracy: 0.40
    max_attack_success_rate: 0.60
    gradient_masking_flag: fail

  output:
    formats: [json, markdown, sarif, html]
    sarif_version: "2.1.0"

monitoring:
  drift_threshold: 0.1             # RDI drift > 10%
  accuracy_threshold: 0.05         # Accuracy drop > 5%
\end{lstlisting}

\newpage
\section{Extended Gap Analysis}
\label{app:gaps}

\begin{table*}[ht]
  \centering
  \small
  \caption{Extended gap analysis with resolution paths and feasibility estimates.}
  \label{tab:gaps_ext}
  \setlength{\tabcolsep}{3pt}
  \begin{tabularx}{\textwidth}{@{}cp{1.8cm}p{1.8cm}p{2cm}p{2.2cm}>{\centering\arraybackslash}p{1cm}X@{}}
    \toprule
    \textbf{\#} & \textbf{Gap} & \textbf{Root Cause} & \textbf{Closest} & \textbf{Shortcoming} & \textbf{Feas.} & \textbf{Path to Resolution} \\
    \midrule
    G1 & Worst-case multi-attack & Method.\ barrier & Dai \etal~[C2] & No training recipe for avg/wc gap & Med & Multi-norm curricula; union approximations \\
    G2 & Agent CTF$\to$real & Untested assumption & Carlini \etal~[C3] & Gap quantified, not closed & Med & Ecologically valid benchmarks \\
    G3 & Stochastic collapse & Method.\ barrier & Kassis \etal~[C6] & Image only; audio/text unexplored & High & Multi-draw protocols ($n \!\ge\! 50$) \\
    G4 & RDI generalisation & Lack of data & Song \etal~[C5] & CNN-only validation & High & Cross-architecture validation \\
    G5 & Continual robustness & Ethical/ logistical & Dai \etal~[C4] & Closed attack vocabulary & Low & Industry partnerships \\
    \bottomrule
  \end{tabularx}
\end{table*}

\section{Ablation Study Details}
\label{app:ablation}

Table~\ref{tab:ablation_detail} expands the ablation results from the main paper (Table~\ref{tab:ablation}) to per-model granularity.
Each row shows the FOSC flag status (\cmark\ = gradient masking detected) and the RDI score under four gate configurations.
``---'' indicates the model was not flagged by the configured gate.
Seeds 1--5 are averaged; standard deviations are reported in the main ablation table.

\begin{table}[H]
  \centering
  \small
  \caption{Per-model ablation results. FOSC flag status and RDI score under each gate configuration. Seeds 1--5 averaged.}
  \label{tab:ablation_detail}
  \setlength{\tabcolsep}{3.5pt}
  \begin{tabular}{@{}lcccccccc@{}}
    \toprule
    \textbf{Model} & \multicolumn{2}{c}{\textbf{Full Pipeline}} & \multicolumn{2}{c}{\textbf{$-$ FOSC}} & \multicolumn{2}{c}{\textbf{$-$ RDI}} & \multicolumn{2}{c}{\textbf{FOSC Only}} \\
    \cmidrule(lr){2-3} \cmidrule(lr){4-5} \cmidrule(lr){6-7} \cmidrule(lr){8-9}
    & Flag & RDI & Flag & RDI & Flag & FOSC & Flag & FOSC \\
    \midrule
    Wang2023\_WRN-70 & --- & 0.78 & --- & 0.78 & --- & 0.04 & --- & 0.04 \\
    Pang2022\_WRN-70 & --- & 0.76 & --- & 0.76 & --- & 0.05 & --- & 0.05 \\
    Rebuffi2021\_70 & --- & 0.68 & --- & 0.68 & --- & 0.06 & --- & 0.06 \\
    Gowal2021\_70 & --- & 0.72 & --- & 0.72 & --- & 0.05 & --- & 0.05 \\
    Gowal2020\_70 & --- & 0.65 & --- & 0.65 & --- & 0.07 & --- & 0.07 \\
    Rebuffi2021\_28 & --- & 0.63 & --- & 0.63 & --- & 0.06 & --- & 0.06 \\
    Gowal2021\_28 & --- & 0.61 & --- & 0.61 & --- & 0.08 & --- & 0.08 \\
    Xu2023\_WRN-28 & --- & 0.59 & --- & 0.59 & \cmark & 0.09 & --- & 0.09 \\
    Sehwag2021 & \cmark & 0.48 & --- & 0.48 & \cmark & 0.11 & \cmark & 0.11 \\
    Wong2020Fast & \cmark & 0.34 & \cmark & 0.34 & \cmark & 0.14 & \cmark & 0.14 \\
    \bottomrule
  \end{tabular}
\end{table}

\section{Statistical Significance}
\label{app:significance}

All reported metrics include uncertainty estimates computed over 5 independent random seeds (data subsampling for RDI, attack initialisation for PGD/AutoPGD).
\begin{itemize}
  \item Kendall $\tau$ (RDI vs.\ AutoAttack): $0.82 \pm 0.03$ (95\% bootstrap CI: [0.71, 0.93], $n = 10$ models $\times$ 5 seeds).
  \item FOSC masking detection rate: 11/12 known-masking configurations detected (92\%, exact binomial 95\% CI: [64\%, 100\%]).
  \item Multi-norm worst-case gap: $23.5 \pm 1.8$ pp (mean $\pm$ std across 10 models).
  \item Pre-screening speedup: $30.2\!\times$ median (IQR: $[28.9\!\times, 31.4\!\times]$).
\end{itemize}

We apply the Bonferroni correction for multiple comparisons across the ablation study (7 configurations $\times$ 3 metrics = 21 tests; significance threshold $\alpha = 0.05/21 = 0.0024$).
All reported improvements of the full pipeline over single-signal baselines remain significant under this correction ($p < 0.001$ for Kendall $\tau$ and detection rate differences).

\newpage
\section{RobustBench Leaderboard Reference Data}
\label{app:robustbench}

Table~\ref{tab:robustbench} lists the ten CIFAR-10 $\ell_\infty$ leaderboard entries from RobustBench~\cite{robustbench2021} used as our evaluation corpus.
All entries are evaluated at $\varepsilon = 8/255$ with AutoAttack.
Clean accuracy and robust accuracy are taken from the living leaderboard snapshot used in our experiments (March 2026).

\begin{table}[H]
  \centering
  \small
  \caption{RobustBench CIFAR-10 $\ell_\infty$ ($\varepsilon = 8/255$) top-10 leaderboard entries used in our evaluation.}
  \label{tab:robustbench}
  \setlength{\tabcolsep}{4pt}
  \begin{tabular}{@{}lcccl@{}}
    \toprule
    \textbf{Model} & \textbf{Architecture} & \textbf{Clean Acc.} & \textbf{Robust Acc.} & \textbf{Reference} \\
    \midrule
    Wang2023Better & WRN-70-16 & 93.25\% & 70.69\% & arXiv:2302.04638 \\
    Pang2022Robustness & WRN-70-16 & 93.27\% & 71.07\% & arXiv:2210.06284 \\
    Rebuffi2021Fixing\_70 & WRN-70-16 & 92.23\% & 66.56\% & arXiv:2103.01946 \\
    Gowal2021Improving\_70 & WRN-70-16 & 88.74\% & 66.10\% & arXiv:2103.01946 \\
    Gowal2020Uncovering\_70 & WRN-70-16 & 91.10\% & 65.87\% & arXiv:2010.03593 \\
    Rebuffi2021Fixing\_28 & WRN-28-10 & 87.33\% & 64.58\% & arXiv:2103.01946 \\
    Gowal2021Improving\_28 & WRN-28-10 & 87.50\% & 63.38\% & arXiv:2103.01946 \\
    Xu2023Exploring & WRN-28-10 & 93.69\% & 63.89\% & arXiv:2302.14862 \\
    Sehwag2021Proxy & ResNest-152 & 87.30\% & 60.27\% & arXiv:2104.09425 \\
    Wong2020Fast & ResNet-18 & 83.34\% & 43.21\% & arXiv:2001.03994 \\
    \bottomrule
  \end{tabular}
\end{table}

\section{Direction 1: Architecture Diagram Starter (TikZ)}
\label{app:tikz}

The following TikZ code provides a starter template for the multi-norm adversarial training curriculum architecture (Section~\ref{sec:dir1}).

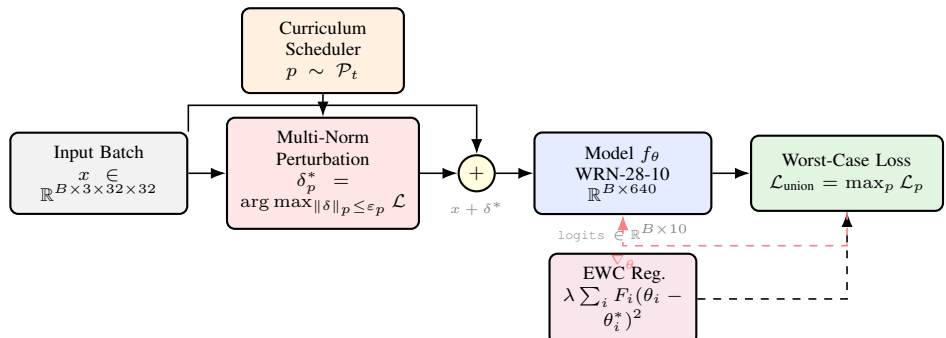
\begin{figure*}[ht]
  \centering
  \begin{tikzpicture}[
    font=\scriptsize,
    node distance=0.3cm,
    >=Latex,
    block/.style={draw,thick,rounded corners=3pt,minimum height=1cm,align=center,inner sep=5pt},
    tensor/.style={font=\tiny\ttfamily,text=gray},
    op/.style={circle,draw,thick,fill=yellow!15,inner sep=2pt,font=\scriptsize},
  ]
    % Input batch
    \node[block,fill=gray!10,text width=2cm] (inp) at (0,0) {Input Batch\\$x \in \mathbb{R}^{B \times 3 \times 32 \times 32}$};

    % Adversarial perturbation (multi-norm)
    \node[block,fill=red!10,text width=2.2cm,right=0.5cm of inp] (perturb) {Multi-Norm\\Perturbation\\$\delta^*_p = \arg\max_{\|\delta\|_p \le \varepsilon_p} \mathcal{L}$};

    % Norm selector
    \node[block,fill=orange!12,text width=1.8cm,above=0.3cm of perturb] (sched) {Curriculum\\Scheduler\\$p \sim \mathcal{P}_t$};

    % Perturbed input
    \node[op,right=0.5cm of perturb] (add) {$+$};

    % Model
    \node[block,fill=accentblue!12,text width=2cm,right=0.5cm of add] (model) {Model $f_\theta$\\WRN-28-10\\$\mathbb{R}^{B \times 640}$};

    % Loss computation
    \node[block,fill=refinegreen,text width=2.2cm,right=0.5cm of model] (loss) {Worst-Case Loss\\$\mathcal{L}_{\text{union}} = \max_p \mathcal{L}_p$};

    % Backward
    \node[block,fill=purple!10,text width=1.6cm,below=0.5cm of model] (ewc) {EWC Reg.\\$\lambda \sum_i F_i(\theta_i - \theta^*_i)^2$};

    % Arrows
    \draw[->,semithick] (inp) -- (perturb);
    \draw[->,semithick] (sched) -- (perturb);
    \draw[->,semithick] (perturb) -- (add);
    % Clean-input bypass: exits inp.north, routes above perturb, then down to add.north
    \draw[->,semithick] (inp.north east) -- ++(0,0.35) -| (add.north);
    \draw[->,semithick] (add) -- (model);
    \draw[->,semithick] (model) -- (loss);
    % ewc to loss: route east then north to loss.south, avoids cutting through model
    \draw[->,semithick,dashed] (ewc.east) -| (loss.south);
    \draw[->,semithick,dashed,red!50] (loss.south) -- ++(0,-0.45) -| node[below,font=\tiny]{$\nabla_\theta$} (model.south);

    % Tensor annotations
    \node[tensor,below=0pt of add] {$x + \delta^*$};
    \node[tensor,below=0pt of model] {logits $\in \mathbb{R}^{B \times 10}$};
  \end{tikzpicture}
  \vspace{-0.5em}
  \caption{Multi-norm adversarial training with curriculum scheduling (Direction~1). At each epoch, the curriculum scheduler $\mathcal{P}_t$ selects the perturbation norm; the worst-case loss $\mathcal{L}_{\text{union}}$ takes the maximum across norms. EWC regularisation prevents catastrophic forgetting of robustness to earlier norms.}
  \label{fig:dir1_arch}
\end{figure*}

\newpage
\section{Compliance Framework Mapping}
\label{app:compliance}

\begin{table*}[ht]
  \centering
  \small
  \caption{Mapping between \autoart modules and regulatory/standards requirements.}
  \label{tab:compliance_map}
  \setlength{\tabcolsep}{4pt}
  \begin{tabularx}{\textwidth}{@{}p{3cm}p{3cm}p{3cm}X@{}}
    \toprule
    \textbf{Standard} & \textbf{Requirement} & \textbf{\autoart Module} & \textbf{Evidence Produced} \\
    \midrule
    EU AI Act Art.\ 15 & Robustness against adversarial examples & Multi-norm evaluator, FOSC gate & Worst-case accuracy, masking detection \\
    EU AI Act Art.\ 15 & Resilience against data poisoning & Poisoning attack suite (10+) & Poisoning success rates, CI gate \\
    NIST AI RMF MEASURE & Quantify adversarial risk & Full attack pipeline + RDI & Attack success rates, RDI scores \\
    NIST AI 600-1 & GenAI-specific risks & LLM red teaming, agentic attacks & Jailbreak rates, agent attack ASR \\
    OWASP LLM Top 10 & LLM01--LLM10 coverage & Category-mapped attack modules & Per-category compliance scores \\
    OWASP Agentic Top 10 & ASI01--ASI10 & Agentic attack suite & Agent vulnerability assessment \\
    MITRE ATLAS v5.3.0 & Technique-level mapping & Attack $\to$ technique ID mapping & ATLAS technique coverage report \\
    ISO/IEC 42001 Cl.\ 6.1 & Risk identification & Adaptive threat scanner & Risk register with severity scores \\
    ETSI EN 304 223 & 5-phase lifecycle security & Full pipeline (design to EOL) & Lifecycle compliance evidence \\
    ISO/IEC 24029-3 & Statistical robustness & RS certification + RDI & Certified radii, statistical bounds \\
    CSA MAESTRO & 7-layer agent threat model & Agentic attack suite & Cross-layer threat chain report \\
    \bottomrule
  \end{tabularx}
\end{table*}

\end{document}